\documentclass{article}

\newlength{\um}
\usepackage{psfig}
\usepackage{epsfig}
\usepackage{amssymb}

\newcommand{\be}{\begin{eqnarray}}
\newcommand{\ee}{\end{eqnarray}}

\newcommand{\ZETF}[3]{{\it Zh.~Eksp.~Teor.~Fiz.} {\bf #1} ({#3}) {#2}}

\newcommand{\PLB}[3]{{\it Phys.~Lett.} {\bf B#1} ({#3}) {#2}}
\newcommand{\YF}[3]{{\it Yad.~Fiz.} {\bf #1} ({#3}) {#2}}
\newcommand{\SJNP}[3]{{\it Sov.~J.~Nucl.~Phys.} {\bf #1} ({#3}) {#2}}
\newcommand{\JETP}[3]{{\it Sov.~Phys.~JETP} {\bf #1} ({#3}) {#2}}
\newcommand{\NPB}[3]{{\it Nucl.~Phys.} {\bf B#1} ({#3}) {#2}}
\newcommand{\PRD}[3]{{\it Phys.~Rev.} {\bf D#1} ({#3}) {#2}}

\newcommand{\ZPC}[3]{{\it Z.~Phys.} {\bf  C#1} ({#3}) {#2}}

\newcommand{\CPC}[3]{{\it Comput.~Phys.~Comm.} {\bf #1} ({#3}) {#2}}

\def\3bar{{\overline 3}}
\begin{document}

\begin{titlepage}

  \begin{flushright}
    LU-TP 97-29\\
    November 1997
  \end{flushright}
\LARGE

\begin{center}
{\bf Investigations into the BFKL Mechanism with a Running QCD Coupling}
\vspace{1cm}

\Large

B. Andersson \footnote{bo@thep.lu.se}, G. Gustafson \footnote{
gosta@thep.lu.se}, H. Kharraziha
\footnote{hamid@thep.lu.se}

\vspace{1cm}
\large

Department of Theoretical Physics\\ 
University of Lund\\ 
S\"{o}lvegatan 14A\\
S-223 62  Lund, Sweden

\end{center}
\normalsize
 
\vspace{1cm}
\begin{abstract} 
 
We present approximations of varying degree of sophistication 
to the integral equations 
for the (gluon) structure functions of
a hadron (``the partonic flux factor'') in a model valid in the
Leading Log Approximation with a running coupling constant. The results are 
all
 of
the BFKL-type, i.e. a power in the Bjorken variable $x_B^{-\lambda}$ with 
the
parameter $\lambda$ determined from the size $\alpha_0$ of the ``effective'' 
running coupling $\bar{\alpha}\equiv 3\alpha_s/\pi=
\alpha_0/\log(k_{\perp}^2)$ and varying
depending upon the treatment of
the transverse momentum pole. We also consider
the implications for the transverse momentum ($k_{\perp}$) fluctuations 
along the emission chains and we obtain an
exponential falloff in the relevant $\kappa\equiv \log(k_{\perp}^2)$-variable, 
i.e.  an inverse power
$(k_{\perp}^2)^{-(2+\lambda)}$ with the same parameter $\lambda$. 
This is different
from the BFKL-result for a fixed coupling, 
where the distributions are Gaussian in the  $\kappa$-variable with a width 
as in a Brownian motion determined by
``the length'' of the emission chains, i.e. $\log(1/x_B)$. The results 
are verified by a realistic Monte Carlo simulation and we provide a simple
physics motivation for the change.
\end{abstract}
\normalsize 
\end{titlepage}

\section{Introduction}
\subsection{The Contents of this Note}

In an earlier paper, \cite{JSBAGG}, we have presented the Linked Dipole Chain
(LDC)
model as a generalization of the well-known CCFM  model (for
Ciafaloni-Catani-Fiorani-Marchesini), \cite{Pinoetal}, to describe the hadronic
structure functions in Deep Inelastic Scattering (DIS) events. 
We have further in
\cite{JSHKBAGG} described  a set of results from the model. 
In this note we would like to
continue the investigations and in particular {\em describe the solutions to 
the equations for the structure functions when a running coupling is 
introduced}. 
We will find power-solutions in the Bjorken-variable, 
i.e. the (gluon) structure 
function behaves as $x_B^{-\lambda}$ in the same way as for 
the BFKL-solutions \cite{BFKL} for a fixed coupling. 

The power $\lambda$ is 
determined by the strength $\alpha_0$ in the effective running coupling
$\bar{\alpha} \equiv 3\alpha_s/\pi=\alpha_0/\ln(k_{\perp}^2)$
(In a pure Yang-Mills theory we have used the value $\alpha_0=12/11$.) and the 
treatment of the
transverse momentum pole, i.e. the small virtualities in our equations.
Actually, it will turn out that the equations become unstable in the sense that
the value of the parameter $\lambda$ is sensitive to the approximate 
treatment of the running coupling in the soft region where $\ln(k_{\perp}^2)$
is small.

If we introduce a
cutoff in the $\ln(k_{\perp}^2)$-variable we obtain, however, inside a large
region, very stable results for an isolated largest eigenvalue $\lambda$. One
major finding is that the LDC model in itself contains a suppression  of the
soft region close to the pole of the coupling constant, and thus 
contributes significantly to this stabilization (This stems
from the averaging over the azimuthal angles).
Our result is then for the value of the isolated largest power $\lambda$ 
that it is about $\lambda \sim (0.3-0.4)\alpha_0$ within a realistic 
cut-off region.

For the transverse 
momentum dependence of the structure function we will find the asymptotic 
solution  $(\ln(k_{\perp}^2))^{\alpha_0/\lambda}$. Further, the transverse 
energy 
distribution along the chain will not, as for the BFKL case with a constant
coupling, be Gaussian in $\log(k_{\perp}^2)$ 
with a width determined, as in a Brownian motion, from the ``length'' 
of the emission chains, i.e. $\log(1/x_B)$ \cite{Alanetal}. 
Instead it will be a (negative) exponential in $\log(k_{\perp}^2)$, i.e. 
it will behave as an inverse power in the transverse momentum
$k_{\perp}^{-2(2+\lambda)}$ with the same $\lambda$-value as above. 
(Note that the
Rutherford parton scattering will behave as $k_{\perp}^{-4}$. The 
extra factor $k_\perp^{-2\lambda}$ is due to the larger $x_B$-values needed
for higher $k_\perp$.) This is also born out by a Monte Carlo simulation.

It is a well-known property of diffusion equations with a force
(in this case stemming from the running coupling which favours small 
$k_\perp$-values) that there are on the short time scale (for 
small $\log(1/x_B)$-values) an erratic behaviour, which can easily be mistaken
for a
stochastic Brownian motion, before the process reaches its long-time stable
distribution. Our estimates of the scales in this case unfortunately indicate
that the HERA range is too small for a noticeable change to the above-mentioned
power behaviour in $k_{\perp}$.

We will end this note with a set of simple examples to show the way the
model interpolates between the DGLAP \cite{DGLAP} and the BFKL mechanisms, 
the reasons
why and the mechanisms by which the stable
distributions in transverse momentum emerge.

\subsection{The CCFM- and the LDC-Models}

\begin{figure}
\psfig{figure=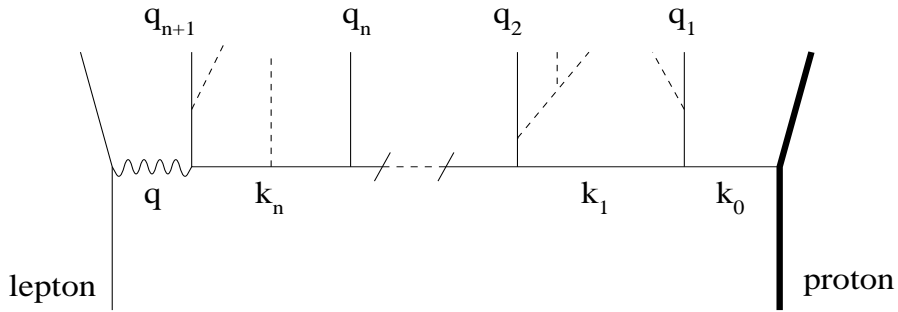,width=12\um,height=4.2\um}
\caption{Fan diagram for a DIS event. $q_i$ denote the emitted ISB partons,
$k_i$ denote the propagators and the dashed lines are the FSB partons.}
\label{fandiagram}
\end{figure}

To introduce the models we note
that it is always possible to subdivide
the radiation in the states into one part, the Initial State Bremsstrahlung
(ISB) (denoted by the vectors $q_j$ in Fig \ref{fandiagram}) and another part,
the Final State Brems\-strahlung (FSB) (the dashed lines in Fig
\ref{fandiagram}). The main requirement is that the FSB partons should be
possible to emit in accordance with the QCD coherence conditions and with
negligible recoils, if the ISB partons are already emitted. The observable
emission weights are given by

\be
\label{ISBweight}
dw= \sum_{I} dw^{(0)}(I) \Delta(I)
\ee
i.e. by the sum over the chosen ISB sets $(I)$ with the basic weight 
$dw^{(0)}(I)$ to emit 
these particular partons and a form factor $\Delta(I)$ stemming from the
radiative corrections from the choice. To be precise each $(I)$-state contains
a sum over all the exclusive states, containing these ISB partons and any other
FSB partons. The FSB should have the property that the corresponding form
factors $Sud(F)$ are of the Sudakov type, meaning that, for each fixed
($I$)-state, the sum over all possible exclusive final states $F$ becomes
unity $\sum_F d\omega_I(F)Sud(F)=1$.

If we ask
for a particular value of the observables $(Q^2,x_B)$, i.e. the (squared) 
``inverse wave-length characteristics of
the probe'' and the scaled energy-momen\-tum, the Bjorken variable $x_B$, then 
we
should in Eq (\ref{ISBweight}) 
sum over the ISB sets, which end on such a configuration.

The choice of the ISB set in the CCFM model is to order all emissions in
rapidity (due to the relation between angle and rapidity this means
angular ordering, a well-known formulation of the QCD coherence 
conditions). Then CCFM chose as ISB the emissions with the property 
that there is no emission further along the rapidity ordering with a 
larger light-cone energy momentum $q_{j+}$. The
ordering is done from the target-hadron side, i.e. the  
target is supposed to have
a (large) positive light-cone energy-momentum $P \equiv (P_+, P_-\simeq 0,
\vec{0}_{\perp})$. In the probe-hadron cms 
the probe (with $-q^2=Q^2$) 
has a large energy-momentum along the negative light-cone 
$q \equiv (-q_+,q_-,\vec{0}_{\perp})$. 
This choice of the ISB is then consistent with QCD
coherence but not with the fact that there should be symmetry between the
hadron and the probe directions, \cite{Yuribook}.

Assuming that the recoils can be neglected, the ISB partons with momenta
$q_j$ are on their mass shells. Energy-momentum conservation at every ISB 
vertex implies that the connector vectors (``the
propagators'') $k_j$, cf. Fig \ref{fandiagram}, fulfill
\be
k_j=P-\sum_1^j q_j & \mbox{ie} & k_j=k_{j-1}-q_j
\ee
CCFM introduce the variables $(z_j,\vec{q}_{\perp j})$ such
that $k_{+j}=z_j k_{+ (j-1)}$ and $\vec{k}_{\perp j}=\vec{k}_{\perp (j-1)}-
\vec{q}_{\perp j}$. In the Leading Log Approximation (LLA), relevant for the 
CCFM
model calculations, the fractional variables should fulfill $z_j \ll 1$, 
i.e. the splitting functions are approximated by $P(z) \sim 1/z$. While the
emitted gluons along the fan diagram in Fig \ref{fandiagram} are treated as
mass less, the connector vectors $k_j$ are space like with $-k_j^2
\simeq \vec{k}_{\perp j}^2 $. This relation is satisfied provided 
\be
k_{\perp j}^2>z_j q_{\perp j}^2
\ee
while smaller $k_\perp$-values are suppressed.
Then the weight and the form factor in Eq (\ref{ISBweight}) is in the CCFM
model
\be
\label{CCFMweight}
\bar{\alpha} \frac{dz}{z}
\frac{dq_{\perp}^2}{q_{\perp}^2}\Delta_{ne}(z,k_{\perp},q_{\perp})
\ee
with the effective coupling $\bar{\alpha}= 3\alpha_s/\pi$ and 
the non-eikonal form factor
\be
\Delta_{ne}= \exp(-\bar{\alpha} \ln(1/z)\ln(k_{\perp}^2/zq_{\perp}^2))
\ee
The LDC model choice in Ref \cite{JSBAGG} is to restrict the ISB emissions in
the CCFM model to those which also fulfill
\be
\label{LDCcriterium}
q_{\perp j}\ge \min(k_{\perp j},k_{\perp (j-1)})
\ee
Gluons which do not satisfy this constraint can be included in the FSB set
causing only small recoils. If the ISB set is restricted in this way,
 and we sum over the corresponding weights and
form factors it is in Ref \cite{JSBAGG} shown that we obtain {\em the same
weight as in Eq (\ref{CCFMweight}) but with the non-eikonal factor 
exchanged for
$1$}. This is obviously a major simplification in order to calculate 
the cross sections according to Eq (\ref{ISBweight}). 

The second simplification comes with respect to the emission of the FSB
radiation. {\em The FSB gluons are emitted as dipole radiation with the
``original'' dipoles spanned between the color-adjacent ISB gluons
$(q_{j},q_{j+1})$ and with the largest allowed FSB transverse momentum
determined by the propagator virtuality $-k_j^2 \simeq k_{\perp j}^2$}. 
This means that we
may for the FSB radiation
make use of the well-known Lund Dipole Cascade model \cite{DIPOLE}, 
 as it is implemented in
the Monte Carlo simulation program ARIADNE, \cite{LL}. 

The transverse momentum restrictions for the FSB emissions, i.e. that the  
allowed FSB transverse momentum is limited by 
the propagator $-k_j^2 \simeq k_{\perp j}^2$,  
occur also in the CCFM model and is a major result from
their very complicated calculations, \cite{Pinoetal}.
Physically it means that the two color adjacent gluon currents from the 
``dipole
pair'' $(q_{j},q_{j+1})$ in the LDC model
do not stem from the same space-time point. There is a distance, which due to
Lorenz contraction is essentially transverse, $b_{\perp} 
\sim 1/\sqrt{-k_j^2}$. The
constraint then stems from the difficulty to emit FSB radiation with a 
wave-length $\lambda \simeq 1/q_{\perp} <b$, i.e. 
smaller than the ``antenna size''. In other words there should be a form factor,
 which if it is e.g. an
inverse power in $k_{\perp}$ is exponentially falling in the relevant
$\ln(k_{\perp}^2)$-variable and therefore leads to negligible contributions 
to the LLA outside this region.

\begin{figure}
\hbox{
\psfig{figure=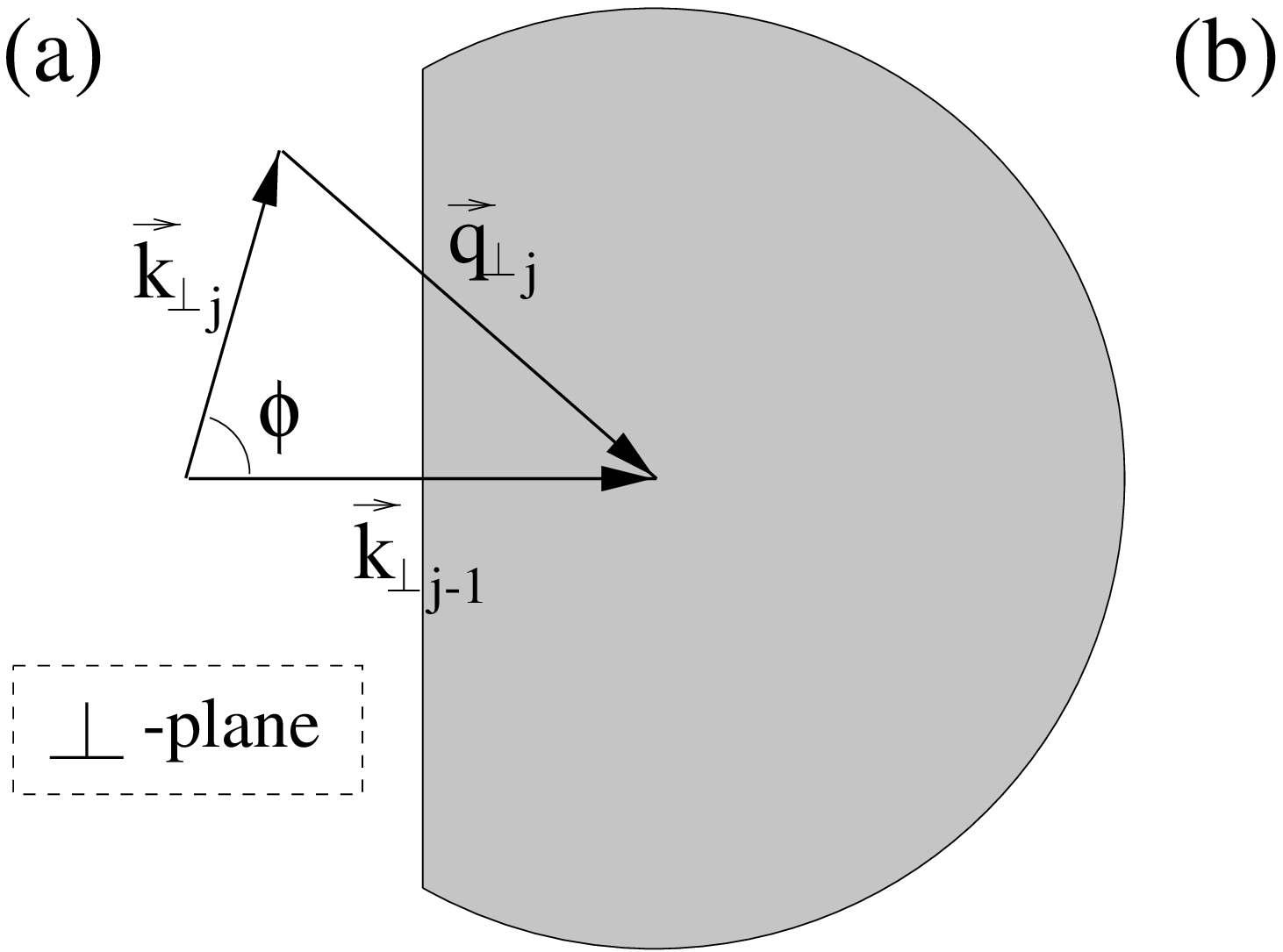,width=6\um,height=4\um}

\psfig{figure=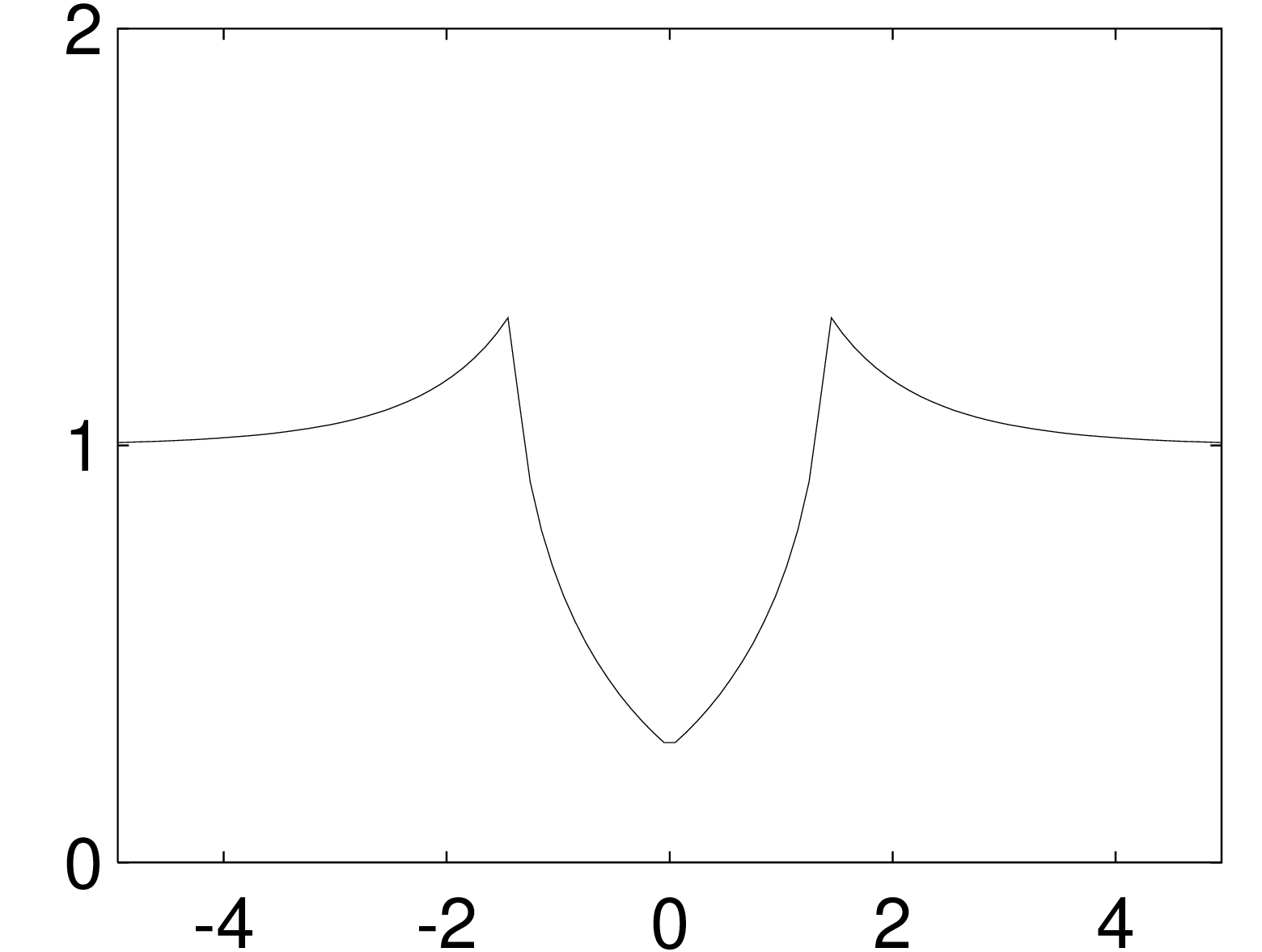,width=6\um,height=4\um}

}
\caption{
(a) The shaded area denotes the forbidden region for $\vec{k}_{\perp j}$
where $q_\perp<\rm{min}(k_{\perp j},k_{\perp j-1})$. 
For $0.5<k_{\perp j}/k_{\perp j-1}<2$, the $\phi$-integral is limited to a
restricted region.
(b) The azimuthal average $h$
as a function of $\ln(k_{\perp j}^2/k_{\perp (j-1)}^2)$.}
\label{azimuth}
\end{figure}

After azimuthal ($\phi$) averaging for fixed $|\vec{k}_{\perp j-1}|$ and
$|\vec{k}_{\perp j}|$ over the transverse pole in 
$\vec{q}_{\perp j}^2 = (\vec{k}_{\perp j}-\vec{k}_{\perp (j-1)})^2$  we obtain
with the constraint in Eq
(\ref{LDCcriterium}) (see Fig \ref{azimuth}a) the LDC weights
\be
\label{LDCweight}
dw(LDC) &=& \bar{\alpha} \frac{dz_j}{z_j} \int \frac{d\phi dk_{\perp j}^2}
{4 \pi \vec{q}_{\perp j}^2} 
\theta\left(q_{\perp j}-{\rm min}\left(k_{\perp j},k_{\perp (j-1)}\right) \right)
\\
\nonumber
&=& 
 \bar{\alpha} \frac{dz_j}{z_j} \frac{dk_{\perp j}^2}{\max(k_{\perp j}^2, 
 k_{\perp (j-1)}^2)} h(k_{\perp j}^2/k_{\perp (j-1)}^2)
\ee
The function $h(a)$, which is obtained from the azimuthal integration, is
given by (see Fig \ref{azimuth}b)

\be
\label{functionha}
h(a) & = &\frac{(1-
\frac{2}{\pi}\arctan\left(\frac{1+\sqrt{a}}{1-\sqrt{a}}
\sqrt{\frac{2\sqrt{a}-1}{2\sqrt{a}+1}}\right)\theta(a-1/4))}{1-a};\ \ \ 0<a<1
\nonumber\\
h(a) & = & h(1/a) 
\ee
As $dz_j/z_j=dy_j$, with $y_j$ equal to the rapidity, we see that the 
expression in Eq. (\ref{LDCweight}) is {\em completely symmetric, and the 
chain
could equally well be generated from the probe end towards the target.}
We also note that in the LLA in general 
$\ln(k_{\perp j}^2/k_{\perp (j-1)}^2)$ is
not close to $0$ and then $h(a) \simeq 1$ (although we note that ``at the
pole'' there is a dip $h(1) =\sqrt{3}/2\pi \simeq 0.28$). Thus in the LLA
approximation (using
$\kappa=\ln(k_{\perp}^2)$ and $\ell= \ln(1/z)$)
\be
\label{LDCweight2}
dw(LDC) \simeq \left\{ \begin{array}{ll}
                      \bar{\alpha}d\ell_jd\kappa_j & \mbox{if $
                      \kappa_j>\kappa_{j-1}$} \\
                      \bar{\alpha} d\ell_j d\kappa_j \exp(\kappa_j
                      -\kappa_{j-1}) & \mbox{otherwise}
                      \end{array}
                 \right.
\ee 
Therefore we obtain in the LDC model from Eqs (\ref{LDCweight}) 
and (\ref{LDCweight2}) 
that {\em the weight distributions  are
local} (i.e. we may define a Markovian stochastic process).
 The locality property in particular implies that 
the process is  simple to implement in a
Monte Carlo generation program which simultaneously describes {\em both the 
structure functions and the final state properties}. Such a general 
program \cite{LDCMC}
is now available and is linked to ARIADNE and JETSET \cite{JETSET}. 

The above-mentioned symmetry implies that we may use either 
the fractional variable
$z_{+j}$ (from the target side) or $z_{-j}$ (from the probe side) as variable
in the Monte Carlo generation. 
In Refs \cite{JSBAGG} 
and \cite{JSHKBAGG} we have discussed the possibility to extend the scenario 
 outside the LLA. The result is that it is necessary to choose the  direction, 
 i.e. to use $z_+$- or
 $z_-$-generation, in accordance with
 the direction of increase  in
transverse momentum, i.e. in the virtuality, of the propagator. Then the
use of the splitting functions in the proper light-cone fraction should contain
many of the non-leading contributions to the structure function. The exact
results for the radiative corrections are, however, not known in this case. 
Nevertheless
the simplicity and symmetry of the result, i.e. the chain of linked dipoles
(containing the FSB and then leading to hadronization in a similar way as in
$e^+e^-$-annihilation events) with
the restriction of no emission above the corresponding propagator virtuality,
seems physically appealing and stable. 

\begin{figure}
\hbox{

\psfig{figure=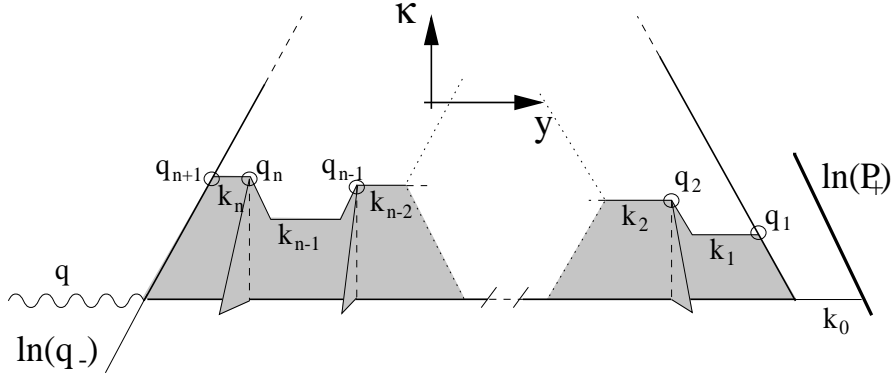,width=12\um,height=5\um}
}
\caption{The Dipole phase space in DIS. The shaded area is the phase space
for FSB.}
\label{triangularfandiagram}
\end{figure}

In Fig \ref{triangularfandiagram} the fan diagram 
emissions in Fig \ref{fandiagram} are described in the well-known Lund dipole 
phase space (which for any dipole is a triangle in $(y,\kappa)$ with 
$y$ the rapidity and with the height and the base-line size given by
$\ln(M^2_{dipole})$). The whole fan emission must occur in between 
the lines corresponding to $\ln(P_+)$ for the target  and $\ln(q_-)$
for the probe
(fixed values of the light-cone energy momenta
$\ln(k_{\pm})=1/2\ln(k_{\perp}^2) \pm y$ correspond to straight lines in the
triangle). The emitted
gluons are described as extended triangular folds, starting at the 
$(y,\kappa)$-value of the
on-shell gluon. The dipoles are spanned between the ``tips'' of these adjacent
gluon folds, i.e. they again correspond to triangles (the two sides 
of the triangular folds correspond to the colour and
anti-colour of the gluon, which are connected each to the two dipoles 
around the
gluon corner). The emission region of each such dipole is limited by 
the maximum virtuality, which
corresponds to the propagator $\ln(-k^2) \simeq \ln(k_{\perp}^2)$.

A propagator in this plot is specified by a horizontal line 
indicating its $\kappa$-value (transverse momentum) and with the 
``starting point'' (right) given by the rapidity value
$y_-\equiv \ln (-k_\perp/k_-)$ and the ``end point'' (left) given by 
$y_+\equiv \ln (k_+/k_\perp)$.
This means that the difference $\delta y $ between the ``endpoint'' 
of one propagator and the ``starting point'' of the next is
\be
\label{logzpm}
0 \leq \delta y  = \ln(k_{\perp max}^2) - \ln(k_{\perp min}^2)
\ee

For completeness we mention that inside the LDC model it is possible, 
Ref \cite{JSBAGG}, to consider different sub-processes in
DIS. Besides the ``ordinary'' parton model description with $Q^2$ exceeding
all the other momentum transfers in the chains, there is the boson-gluon fusion
process, where the the last splitting exceeds $Q^2$ and all the other ones.
Finally in a Rutherford parton scattering process the largest momentum 
transfer, $k_{\perp max}^2$, will occur somewhere in the middle of the chain. 
Then we may consider the result of
the dipole chain generation as the convolution of the target and 
the probe structure functions (each developed to the virtuality $-k_{max}^2$
with the Bjorken variables $x_{\pm}$, respectively) and multiplied by the
well-known Rutherford transverse momentum pole $(k_{max}^2)^{-2}$. This also
implies that such a Rutherford scattering $k_{\perp}$ must
always be larger than all bremsstrahlung $k_{\perp}$ in the event. This
constraint
produces an effective cut-off for small $k_{\perp}$ Rutherford scattering.

\section{The LDC Evolution Equations}
\subsection{Integral Equations}

We study in this paper only purely gluonic chains, and we let the 
gluonic structure function $F(x,Q^2)$ denote the gluon density 
in $\ln(1/x)$. (Thus $F/x$ is the density in $x$.) 
In this section we will treat the LDC evolution equation for the
``non-integrated'' structure function, 
${\cal F}(x,k_\perp^2)$ defined by

\begin{equation}
\label{defcalF}
F(x,Q^2)=
\int_{k_{\perp cut}^2}^{Q^2}
\frac{dk_\perp^2}{k_\perp^2}{\cal F}(x,k_\perp^2)
+
\int_{Q^2}^{Q^2\!/x}
\frac{dk_\perp^2}{k_\perp^2}\frac{Q^2}{k_\perp^2}{\cal 
F}(x\frac{k_\perp^2}{Q^2},k_\perp^2).
\end{equation}
Here $k_\perp$ is the transverse momentum of the last 
link in the chain. We note that the contribution from 
chains with $k_\perp^2>Q^2$ is suppressed by the factor 
$Q^2/k_\perp^2$, which makes it sub-leading. We also note 
the modified argument, $x \rightarrow x \cdot k_\perp^2/Q^2$, 
obtained for this term within the LDC model. Although less 
important this modification further suppresses this term.

 From the weights in Eq (\ref{LDCweight}) we find a 
 recursive relation for ${\cal F}$, and in \cite{JSBAGG} 
 we derived the following evolution equation:

\be
\label{mastercalF}
\frac{\partial {\cal F}}{\partial \ell}(\ell,\kappa) &=& 
\int_{\kappa_{cut}}^{\kappa} d\kappa^{\prime}   \bar{\alpha}
(\kappa) h(\kappa-\kappa^{\prime}) {\cal F}(\ell,\kappa^{\prime})
\nonumber\\ 
&+& \int_{\kappa}d\kappa^{\prime}  \bar{\alpha}(\kappa^{\prime})  
h(\kappa^{\prime}-\kappa) \exp[-(\kappa^{\prime}-\kappa)]
{\cal F}(\ell +\kappa-\kappa^{\prime},\kappa^{\prime})
\ee
We have here used the variables $\ell=\ln(1/x)$ and 
 $\kappa=\ln(k_\perp^2)$, introduced in the previous section. 
 The function $h$ defined in Eq (\ref{functionha}) originates 
 from the azimuthal integration in Eq (\ref{LDCweight}). 
 From now on we express $h$ in the logarithmic variables 
 $\kappa$ and $\kappa^{\prime}$, and note that it depends 
 only on the difference $\mid \!\!\kappa - \kappa^{\prime} \!\! \mid$. 
 (Note that in the LLA we can put $h=1$. The argument 
 of ${\cal F}$  in the second integral is explained in Eq (\ref{logzpm})). 
 
 The first integral in Eq (\ref{mastercalF}) corresponds to chains, 
 where the last step is upwards in transverse momentum 
 from $\kappa^{\prime}$ to $\kappa$, and the second integral, 
 where $\kappa^{\prime}$ is larger than $\kappa$, to chains 
 with a final step downwards. For the scale in 
the running coupling we have taken the largest virtuality 
in each vertex. Thus in the first integral we have 
$\bar{\alpha}(\kappa)$, while in the second term we have
$\bar{\alpha}(\kappa^{\prime})$. 

With a running $\alpha_s$ the result will (as we will see in 
the following) be rather sensitive to the behaviour
in the soft region for small $k_\perp$. Naturally a perturbative
calculation cannot be trusted when $\alpha_s$ becomes very large. 
Lacking
a good understanding of the soft region, some kind of cut-off 
is necessary, and in this paper we have studied the results
obtained assuming a lower cut-off, $k_{\perp cut}$ or 
$\kappa_{cut}=\ln(k_{\perp cut}^2/\Lambda^2)$, in Eqs 
(\ref{defcalF}) and (\ref{mastercalF}), while keeping the
form ${\bar \alpha}=\alpha_0/\kappa$ for the coupling constant. 
(Another possibility would be to study the result assuming that 
$\alpha_s$ saturates
for $k_\perp$-values below some cut.)

After a Mellin
transform in $x_B$, which corresponds to a Laplace transform 
in $\ell$ with
\be
\label{definef}
f_{\lambda}(\kappa)= \int d\ell \exp(-\lambda \ell) 
{\cal F}(\ell, \kappa)
\ee
we obtain (we will in general neglect the index $\lambda$ on
$f$ from now on)
\newpage
\be
\label{masterf}
\lambda f(\kappa)&=& \int^{\kappa} d\kappa^{\prime} \bar{\alpha}(\kappa) 
h(\kappa-\kappa^{\prime}) 
f(\kappa^{\prime}) \nonumber \\ &+& \int_{\kappa} d\kappa^{\prime} 
\bar{\alpha}(\kappa^{\prime}) h(\kappa^{\prime}-\kappa)  \exp[-(\lambda 
+1)(\kappa^{\prime}-\kappa)]
f(\kappa^{\prime})
\ee
It is sometimes useful (cf section \ref{dominpath}) to introduce a symmetric 
non-integrated
structure function ${\cal F}_s$ by the definition
\be
\label{symmetriccalF}
{\cal F}_s(\ell,\kappa)={\cal F}(\ell,\kappa)\exp(-\kappa/2)
\ee
In that way the exponential suppression factor
$\exp-(\kappa^{\prime}-\kappa)$ in the second integral of Eq
(\ref{mastercalF}), which is ``paid'' only ``for going down'' in
$\kappa$, is changed so that ``we pay the same amount for going up as
for going down'' in $\kappa$. 

\subsection{The Case of a Fixed Coupling $\bar{\alpha}$}
\label{subfixed}

The case of a fixed coupling has been treated in Ref \cite{JSBAGG} 
(and also at
many other places)  
and we will only briefly describe (some of) the
results for future reference. We perform a Laplace transform 
with respect to
$\kappa$, and define $\hat{f}$ by
\be
\label{definehatf}
\hat{f}(\gamma)=\int d\kappa \exp(-\gamma \kappa) f(\kappa)
\ee
As $h$ is a function of
$\kappa - \kappa^{\prime}$, we actually work with a convolution
integral, and pole-singularities are obtained in $\hat{f}$ at 
the $\gamma$-values
(the ``anomalous dimensions'' for given $\lambda$) which satisfy
\be
\label{masterfhat}
\lambda= \bar{\alpha} [g(\gamma)+g(\lambda+1-\gamma)] & \mbox{with} &
g(\gamma)= \int^{\infty}_0 d\Delta \exp(-\gamma \Delta) h(\Delta)
\ee
The two $g$-contributions stem from the two integrals  in
Eq (\ref{masterf}), respectively. and $h(\Delta)$ is here 
interpreted as a function of the logarithmic variable 
$\Delta = \kappa - \kappa^{\prime}$.
 
There are different levels of approximations possible to apply 
in the treatment of the Eqs
(\ref{mastercalF}), (\ref{masterf}) and (\ref{masterfhat}). 
 The simplest one is to
make use of the approximation in Eq (\ref{LDCweight2}) 
and to replace $h$ by 1, which gives $g \approx 1/\gamma$. The function $h$ deviates from 1 around the singular point
 $\ln(\kappa_j^2/\kappa_{j-1}^2) = 0$, and in an improved treatment we can describe this deviation by a $\delta$-subtraction 
(cf ref \cite{JSBAGG}):
\be
\label{delta}
h\simeq \hat{h} = [1 - A
\delta(\kappa - \kappa^{\prime})]
\ee
Inserted into  Eq (\ref{masterfhat}) this gives

\be
\label{approx1A}
g(\gamma)\simeq\bar{\alpha} \,[1/\gamma-A/2]\,\Rightarrow\, 
\lambda \simeq \bar{\alpha}\, [\frac{1}{\gamma} + 
\frac{1}{\lambda +1 -\gamma} - A]
\ee
If we fix $A$ by the condition 
$\int(\hat{h}(\Delta)-1)d\Delta = \int(h(\Delta)-1)d\Delta$ 
we find $A\approx 0.65$. In $g$ also other moments 
of $h$ are important, and in ref \cite{JSBAGG} 
it was found that for $A\approx 0.8$ the same 
$\lambda$-value was obtained from the original $h$ and the 
approximation $\hat{h}$. 

We note that the BFKL result corresponds to a neglect of the
$\lambda$-dependence in the second $g$-term in Eq (\ref{masterfhat}). 
Although this is a
sub-leading term, it is numerically not small, 
as discussed in Ref \cite{JSBAGG}. 
(There is also in BFKL a different method to 
regularize the singularity at  $\ln(\kappa_j/\kappa_{j-1}) \sim 0$.) 

In \cite{JSBAGG} we also pointed out another important non-leading 
effect. For consistency it is necessary that
``the steps'' in $\ln(1/z)$ really are large, but it turns out 
that there are
essential
contributions to the BFKL-integrals from eg $z>1/2$. 
We return to this problem in section \ref{xdep}. 

 The $x_B$-dependence is then determined by the inverse Mellin 
(or Laplace)
transform. The leading behaviour stems from a pinch singularity 
at $\lambda =
\lambda_s$, 
when the two symmetrical solutions to Eq (\ref{masterfhat}) 
coincide, i.e. when
$\gamma=(1+\lambda_s)/2$ (or $\gamma=1/2$ if the $\lambda$-dependence 
in $g(\lambda+1-\gamma)$ is neglected.) From Eq (\ref{approx1A}) 
we find e.g. for $\bar{\alpha}=0.2$ and $A=0.8$ the result 
$\lambda_s \approx 0.41$ (neglecting the $\lambda$-dependence 
in $g$, we would instead obtain $\lambda_s \approx 0.64$), with the
corresponding low-$x_B$ behaviour $\sim$$x_B^{-\lambda_s}$.

When the inverse Laplace(Mellin)-transform is performed one 
also obtains
the dependence on $k_{\perp}$ (or rather on 
$\kappa=\ln(k_{\perp}^2)$) by a
saddle-point approximation around the pinch singularity. The
result (besides essentially trivial kinematic factors) is a 
Gaussian in
$\kappa$ with a width $\sim$$\sqrt{\ln(1/x})$, a result 
which can be interpreted as a kind of random walk in $\kappa$-space
\cite{Alanetal}.

\subsection{The Case of a Running Coupling 
$\bar\protect{\alpha\protect}$}

For a running coupling, chains with larger $\kappa$-values are 
disfavoured by the smallness of $\alpha_s$. This will imply that 
$<\!\kappa\!>$ does not increase beyond any bound as in the case with 
a fixed coupling. Instead the $\kappa$-distribution saturates, and
 for small $x$, ${\cal F}(x,k_\perp^2)$
approaches asymptotically a factorizing form

\begin{equation}
\label{factform}
{\cal F}(x,\kappa)=x^{-\lambda}f(\kappa).
\end{equation}

 The result appears to
be similar to the type of random walk an atmospheric molecule follows 
in the
earth's gravitational field. In this case, when the molecules 
are constantly
pulled downwards, an equilibrium is obtained for an exponentially 
decreasing
density (see also the discussion in Appendix \ref{randomwalk}). 
As discussed 
in section \ref{ktdep}, also in our model with 
a running $\alpha_s$
the asymptotic $k_\perp$-distribution is an exponential in 
$\kappa = \ln k_\perp^2$. Inserting the asymptotic form in Eq 
(\ref{factform}) into Eq (\ref{mastercalF}) implies that 
$f(\kappa)$ must satisfy Eq (\ref{masterf}) above, 
now with $\bar\protect {\alpha\protect}(\kappa)=\alpha_0/\kappa$, 
where $\alpha_0=12/11$ if we study a purely gluonic situation 
with no quarks. The solution to this equation will be studied 
in the following section.

\section{The Solution for Running Coupling}
\label{running}

We have not been able to find an exact analytic solution to the 
integral equation (\ref{masterf}). Instead we have solved 
it numerically, using Chebyshev polynomials to convert it 
to a matrix equation. This method was also used in ref \cite{JanChebyshev}, 
and is discussed further 
in appendix \ref{numapp}. The
$x$-dependence for small $x$-values is determined by the largest
eigenvalue $\lambda$ to this matrix equation.

In order to better understand the properties of the model 
we have also studied a set of
approximations, which all agree with the original 
Eq (\ref{masterf}) to leading logarithmic accuracy:

\begin{itemize}
{\item[a)] by using $\kappa$ instead of $\kappa'$ for the 
argument of
$\bar\alpha$ in the second integral of Eq (\ref{masterf})} 
{\item[b)] by using the approximation 
$h \approx (1 - A\delta(\kappa - \kappa^{\prime}))$ 
in Eq (\ref{delta}).
 A special case (called b') with $A=0$ corresponds 
 to $h \rightarrow 1$.}
{\item[c)] with both of the approximations a) and b) 
(cf section \ref{quaprop}).
(Case c') corresponds to both approximations a) and b').)}
\end{itemize}

The cases b') and c) can be studied analytically. Case b') is
discussed in appendix \ref{solution}, and the simpler case 
c) will be discussed in section \ref{quaprop}. 

\subsection{Qualitative properties}
\label{quaprop}

In order to understand the qualitative features of Eq 
(\ref{masterf}) we start
by investigating the approximation called c) above, 
which is obtained when 
${\bar \alpha}(\kappa')$ is replaced by
${\bar \alpha}(\kappa)$ in the last integral of Eq (\ref{masterf}), 
and the variation of $h$ is approximated by a $\delta$-function 
according to Eq (\ref{delta}).
We note that both these approximations are of non-leading order, 
and the solution has properties similar to the solution of 
the original Eq (\ref{masterf}). With these approximations 
the integral equation in Eq (\ref{masterf}) 
can be transformed by straightforward
differentiations into the second-order differential equation
\be
\label{BSapprox}
u\frac{d^2f}{du^2}=[(\lambda+1)u-2] \frac{df}{du}+(\lambda+1)(1-\mu)f
\ee
where $u=\kappa + \mu A$ and $\mu=\alpha_0/\lambda$. This is 
equivalent to (the
radial part of) the Schroedinger equation for the hydrogen atom 
with the angular momentum variable
$\ell=0$. To see that, we introduce the function $\psi$ defined by
\be
\label{fdefinepsi}       
f= \exp[(\lambda+1)/2)u] \psi(u)
\ee
We obtain by straightforward calculations
\be
\label{Schroedingerhydrogen}
-\frac{1}{2}\frac{d^2 (u\psi)}{udu^2} -\frac{\mu(\lambda+1)\psi}{2u}=
-\frac{(\lambda+1)^2\psi}{8}
\ee
Thus the ``mass''-value is unity and the (squared) ``charge'' equal to
$\mu(\lambda+1)/2$.

For large values of $\kappa$ (large
values of $u$) the second term in Eq (\ref{Schroedingerhydrogen}) 
can be neglected,
and the solution behaves as $u\psi\sim
\exp [\pm(1+\lambda)\kappa/2]$. Here the physical solution must
correspond to the minus sign. From Eq (\ref{fdefinepsi}) 
this implies that $f$
varies as a power of $\kappa$ for large $\kappa$-values. 
From the behaviour
of eq. (\ref{BSapprox}) for large $u$, ie large $\kappa$, 
we find 
\be
\label{fasymptot}
f\propto \kappa^{\frac{\alpha_0}{\lambda}-1};\,\,\,\,\kappa \ \mathrm{large}
\ee

The value of $\lambda$ is determined by the boundary condition 
for small $\kappa$, which can be derived from the integral equation 
(\ref{masterf}). (With the present approximation we find
$f'(\kappa_{cut})=[1+\lambda - 
1/(\kappa_{cut}+\mu A)]f(\kappa_{cut})$.)
This implies that we have solutions for a discrete
set of eigenvalues $\lambda$. The asymptotic behaviour
of the solution is determined by the largest of these eigenvalues, 
which
corresponds to a solution $\psi$ with no zeros between
$\kappa_{cut}$ and $\infty$.

 We note that since $\lambda$ is fixed by the boundary condition, 
 the result is sensitive to the value of
$\kappa_{cut}$, i e it is sensitive to the soft region. 
When $\kappa_{cut}$
increases, the value of $\lambda$ decreases continuously. 
Note, however, that if we include the correction term in $h$ 
then this sensitivity to $\kappa_{cut}$ is strongly reduced. 
As $\kappa =\kappa_{cut}$ implies $u_{cut}= \kappa_{cut} +\mu A$ 
we keep away from the singular point $u=0$ in Eq 
(\ref{Schroedingerhydrogen}), even when  
$\kappa_{cut}$ approaches 0. This stabilizing effect of 
$h$ will be further discussed in the following subsection.

For a {\em fixed} coupling we can from Eq (\ref{masterf}) 
derive a differential equation similar to Eq (\ref{BSapprox}). 
This equation does not, however, contain the factors $u$ 
in front of $d^2 f/du^2$ and $df/du$. 
Therefore this equation has two exponentially growing solutions, 
which implies that in this case the dependence of 
${\cal F}$ upon $x$ and $\kappa$ does not factorize.

Summarizing we have found that the asymptotic behaviour of 
${\cal F}$ is given by the form 
\be
\label{asymptotcalF}
{\cal F} \propto x^{-\lambda}\cdot \kappa^{\frac{\alpha_0}{\lambda} -1} \equiv 
x^{-\lambda}\cdot (\ln k_{\perp}^2)^{\frac{\alpha_0}{\lambda} -1} ; 
\,\,\,\,\ln(1/x) \gg\kappa\gg1
\ee
where the power $\lambda$ is an eigenvalue determined by the boundary 
conditions. It turns out that these qualitative features do not 
rely 
on the approximations used in this subsection, 
but are also relevant for the solution to the original equation 
(\ref{masterf}).

\subsection{The $x$-dependence} 
\label{xdep}

We will now study the numerical solution to Eq (\ref{masterf}).
As the second integral is sub-leading, we see that for
${\bar \alpha}=\alpha_0/\kappa$ the eigenvalue $\lambda$ will
to first approximation be proportional to $\alpha_0$. Since we
are here studying only purely gluonic chains, and are not including
quark links, it may be most consistent to use a value of $\alpha_0$
which corresponds to a pure Yang-Mills theory. From
the relation $\alpha_0=36/(33-2n_f)$ we find for a pure Yang-Mills
$\alpha_0=12/11$ as compared to e.g. $\alpha_0=4/3$ for $n_f=3$.

\begin{figure}
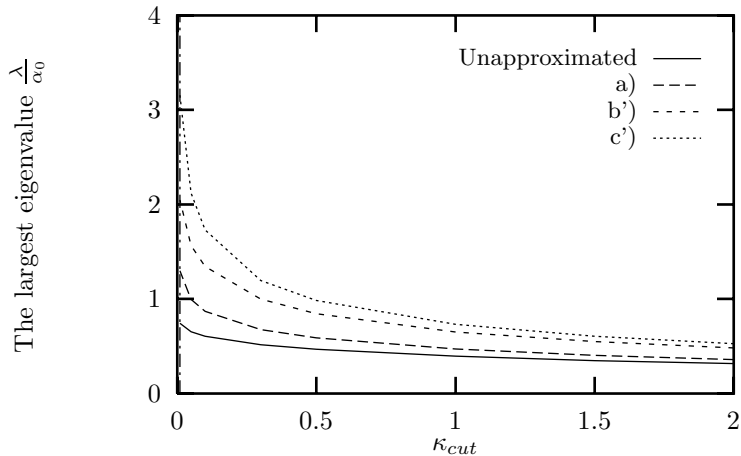


\input lgcomp.tex
\caption{The dependence of $\lambda/\alpha_0$ on $\kappa_{cut}$. The solid line
is from eq. (\ref{masterf}) without approximations. The other lines are
for the approximations a, b' and c' from down and up. As the calculations
do not include quarks we have presented results for the more stable
ratio $\lambda/\alpha_0$, where $\alpha_0=12/11$ for $n_f=0$ and
$\alpha_0=4/3$ for $n_f=3$.}
\label{lgcomp}
\end{figure}

As we have not studied the influence from the quarks we present
results for the more stable quantity $\lambda/\alpha_0$. Thus
in Fig (\ref{lgcomp}) we show how $\lambda/\alpha_0$ depends on the
value of $\kappa_{cut}$ 
in the range $0.01<\kappa_{cut}<2$ (To be exact the results presented 
correspond to $\alpha_0=12/11$ and for $\alpha_0=4/3$ they are about 4\%
smaller). Results are presented 
for the original equation (\ref{masterf}) and also for 
the approximations a), b') and c') defined above. 
We see that the result is sensitive to both the soft cut-off, 
$\kappa_{cut}$, and to the non-leading modifications in the 
different approximations. We note, however, that the factor 
$h$ in the
LDC model provides a very strong stabilizing effect. 
Thus for the original non-approximated equation, the result 
is fairly stable
in a range $0.5<\kappa_{cut}<2.0$, which ought to contain 
realistic values for
$\kappa_{cut}$. In this range we find 
$0.3\, \alpha_0<\lambda <0.5\, \alpha_0$.

In the LLA the energy fractions $z_i$ are assumed to be small 
compared to 1, and in ref \cite{JSBAGG} we pointed out
 that quantitatively significant contributions arise from 
 large $z$-values. We here show that this is true also
  with a running coupling.
For the splitting function, we have in the weights 
in Eq (\ref{LDCweight}) used $P(z)\propto 1/z$, which
is correct within the leading log approximation. 
The full splitting function without regularization for $z=1$ reads

\begin{equation}
\label{fullsplit}
P(z)=2N_c\left\{ \frac{1-z}{z}+\frac{z}{1-z}+z(1-z)\right\}
\end{equation}
If inserted in an evolution equation, the pole at $z=1$ has 
to be regularized.
Virtual corrections imply that this pole does not contribute 
to the increase of the cross section (ie the structure functions) 
for increasing $Q^2$. One way to interpret this is to note 
that in the gluon splitting process the ``old''
gluon disappears and is replaced by two ``new'' gluons. 
Thus only one new extra
gluon is produced, and values of $z$ close to 
the pole $1/(1-z)$ corresponds to the original gluon
losing some of its energy. Thus this contribution does not 
increase the cross section, but only ``shifts it'' to slightly 
smaller $x$-values.

\begin{figure}
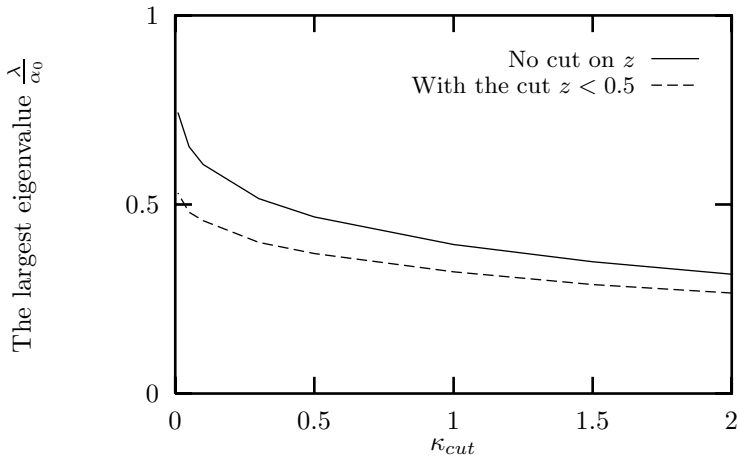


\input twoeff.tex
\caption{The dependence of $\lambda/\alpha_0$ on $\kappa_{cut}$. 
The solid line 
is for the default splitting function. The dashed line is for $P(z>0.5)=0$.
As in Fig (\ref{lgcomp}) we present results for the more stable ratio 
$\lambda/\alpha_0$.}
\label{twoeff}
\end{figure}

In the LLA the energy fraction $z$, which corresponds to the 
``new'' gluon, is assumed to be
small. If we want to reduce recoils as much as possible, 
we can say that the
``new'' gluon by definition is the one with least energy, 
which must imply
that $z<0.5$ (c.f. ref.). Therefore it might be sensible 
to replace the factor $1/z$ in the weight in Eq (\ref{LDCweight}) 
by $(1/2N_c) P(z) \theta(0.5 - z)$. To get some estimate of this 
effect we note that $P(z)/2N_c$ is smaller than $1/z$ in most 
of the region $0 < z < 0.5$. (Actually $P(z)/2N_c - 1/z$ is 
positive only for $z > 0.43$, and also here it is fairly small.) 
Therefore to estimate qualitatively the consequence of this 
non-leading effect, we
have studied the changes obtained from the replacement 
$1/z \rightarrow 1/z\cdot\theta(0.5-z)$. It is straightforward 
to show that this change will give an extra factor $2^{\lambda}$ 
to the left hand side of Eq (\ref{masterf}), which means that 
$\lambda f$ 
will be replaced by $2^{\lambda} \lambda f$.
Apparently, this leads to smaller values of $\lambda$ with an 
approximate
factor $2^{-\lambda}\approx 1/\sqrt{2}$. 
In Fig \ref{twoeff}, we compare 
$\lambda(\kappa_{cut})$ for the two alternative splitting 
functions. This
non-leading correction is clearly significant, reducing $\lambda$ by
20-25\%.

\begin{figure}
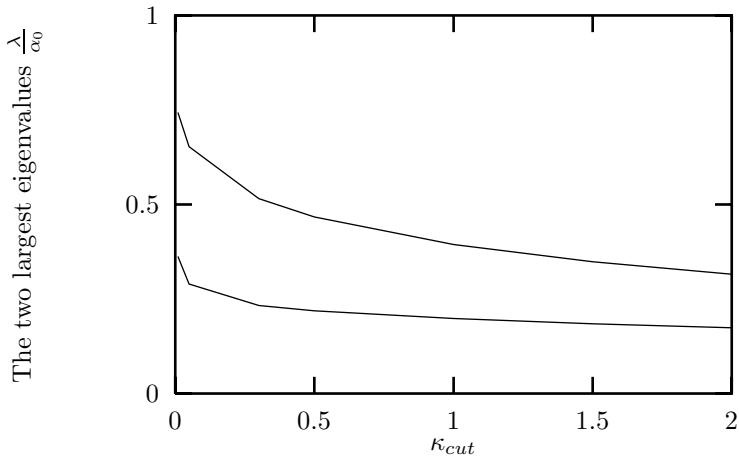


\input lambda00.tex
\caption{The two largest eigenvalues, as a function of $\kappa_{cut}$}
\label{l00}
\end{figure}

We have also checked that the largest eigenvalue is well separated from the smaller ones (see Fig \ref{l00}). Thus for $\kappa_{cut} = 1$ we find for the largest and the second largest eigenvalue the values 0.4 $\alpha_0$ and 0.2 $\alpha_0$. Thus for $\ln(1/x) \gtrsim 5$ the non-leading contribution should be supressed by a factor $e^{-1}$ (and for $\ln(1/x) \gtrsim 10$ by a factor $e^{-2}$).

\subsection{Transverse Momentum Dependence}
\label{ktdep}
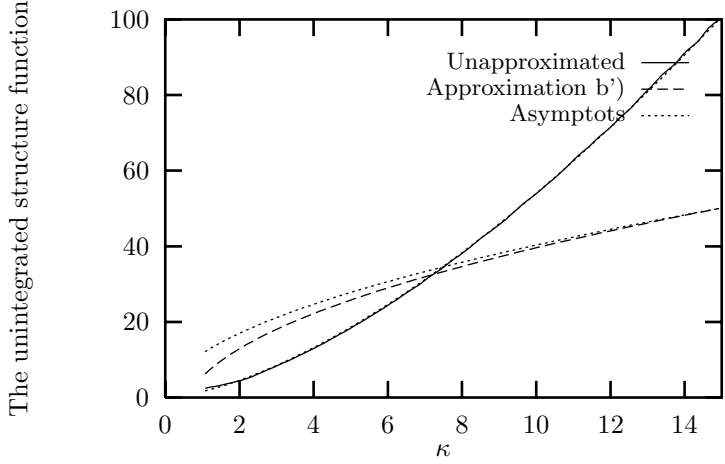
\begin{figure}

\setlength{\unitlength}{0.1bp}
\special{!
/gnudict 40 dict def
gnudict begin
/Color false def
/Solid false def
/gnulinewidth 5.000 def
/vshift -33 def
/dl {10 mul} def
/hpt 31.5 def
/vpt 31.5 def
/M {moveto} bind def
/L {lineto} bind def
/R {rmoveto} bind def
/V {rlineto} bind def
/vpt2 vpt 2 mul def
/hpt2 hpt 2 mul def
/Lshow { currentpoint stroke M
  0 vshift R show } def
/Rshow { currentpoint stroke M
  dup stringwidth pop neg vshift R show } def
/Cshow { currentpoint stroke M
  dup stringwidth pop -2 div vshift R show } def
/DL { Color {setrgbcolor Solid {pop []} if 0 setdash }
 {pop pop pop Solid {pop []} if 0 setdash} ifelse } def
/BL { stroke gnulinewidth 2 mul setlinewidth } def
/AL { stroke gnulinewidth 2 div setlinewidth } def
/PL { stroke gnulinewidth setlinewidth } def
/LTb { BL [] 0 0 0 DL } def
/LTa { AL [1 dl 2 dl] 0 setdash 0 0 0 setrgbcolor } def
/LT0 { PL [] 0 1 0 DL } def
/LT1 { PL [4 dl 2 dl] 0 0 1 DL } def
/LT2 { PL [2 dl 3 dl] 1 0 0 DL } def
/LT3 { PL [1 dl 1.5 dl] 1 0 1 DL } def
/LT4 { PL [5 dl 2 dl 1 dl 2 dl] 0 1 1 DL } def
/LT5 { PL [4 dl 3 dl 1 dl 3 dl] 1 1 0 DL } def
/LT6 { PL [2 dl 2 dl 2 dl 4 dl] 0 0 0 DL } def
/LT7 { PL [2 dl 2 dl 2 dl 2 dl 2 dl 4 dl] 1 0.3 0 DL } def
/LT8 { PL [2 dl 2 dl 2 dl 2 dl 2 dl 2 dl 2 dl 4 dl] 0.5 0.5 0.5 DL } def
/P { stroke [] 0 setdash
  currentlinewidth 2 div sub M
  0 currentlinewidth V stroke } def
/D { stroke [] 0 setdash 2 copy vpt add M
  hpt neg vpt neg V hpt vpt neg V
  hpt vpt V hpt neg vpt V closepath stroke
  P } def
/A { stroke [] 0 setdash vpt sub M 0 vpt2 V
  currentpoint stroke M
  hpt neg vpt neg R hpt2 0 V stroke
  } def
/B { stroke [] 0 setdash 2 copy exch hpt sub exch vpt add M
  0 vpt2 neg V hpt2 0 V 0 vpt2 V
  hpt2 neg 0 V closepath stroke
  P } def
/C { stroke [] 0 setdash exch hpt sub exch vpt add M
  hpt2 vpt2 neg V currentpoint stroke M
  hpt2 neg 0 R hpt2 vpt2 V stroke } def
/T { stroke [] 0 setdash 2 copy vpt 1.12 mul add M
  hpt neg vpt -1.62 mul V
  hpt 2 mul 0 V
  hpt neg vpt 1.62 mul V closepath stroke
  P  } def
/S { 2 copy A C} def
end
}
\begin{picture}(2880,1728)(0,0)
\special{"
gnudict begin
gsave
50 50 translate
0.100 0.100 scale
0 setgray
/Helvetica findfont 100 scalefont setfont
newpath
-500.000000 -500.000000 translate
LTa
LTb
600 251 M
63 0 V
2034 0 R
-63 0 V
600 536 M
63 0 V
2034 0 R
-63 0 V
600 821 M
63 0 V
2034 0 R
-63 0 V
600 1107 M
63 0 V
2034 0 R
-63 0 V
600 1392 M
63 0 V
2034 0 R
-63 0 V
600 1677 M
63 0 V
2034 0 R
-63 0 V
600 251 M
0 63 V
0 1363 R
0 -63 V
880 251 M
0 63 V
0 1363 R
0 -63 V
1159 251 M
0 63 V
0 1363 R
0 -63 V
1439 251 M
0 63 V
0 1363 R
0 -63 V
1718 251 M
0 63 V
0 1363 R
0 -63 V
1998 251 M
0 63 V
0 1363 R
0 -63 V
2278 251 M
0 63 V
0 1363 R
0 -63 V
2557 251 M
0 63 V
0 1363 R
0 -63 V
600 251 M
2097 0 V
0 1426 V
-2097 0 V
600 251 L
LT0
2394 1514 M
180 0 V
750 286 M
19 4 V
20 3 V
19 4 V
20 5 V
19 4 V
20 5 V
20 5 V
19 6 V
20 7 V
19 9 V
20 8 V
19 9 V
20 8 V
20 9 V
19 9 V
20 8 V
19 9 V
20 10 V
19 9 V
20 10 V
20 10 V
19 10 V
20 10 V
19 11 V
20 11 V
19 11 V
20 10 V
20 12 V
19 12 V
20 11 V
19 12 V
20 12 V
19 12 V
20 12 V
20 13 V
19 13 V
20 12 V
19 14 V
20 13 V
19 13 V
20 12 V
20 15 V
19 15 V
20 13 V
19 13 V
20 14 V
19 16 V
20 15 V
20 13 V
19 15 V
20 15 V
19 16 V
20 17 V
19 14 V
20 14 V
20 16 V
19 14 V
20 16 V
19 18 V
20 17 V
19 18 V
20 16 V
20 15 V
19 15 V
20 17 V
19 17 V
20 16 V
19 17 V
20 19 V
20 18 V
19 20 V
20 19 V
19 16 V
20 16 V
19 19 V
20 19 V
20 16 V
19 17 V
20 19 V
19 20 V
20 17 V
19 17 V
20 22 V
20 22 V
19 19 V
20 20 V
19 21 V
20 22 V
19 16 V
20 16 V
20 20 V
19 23 V
20 20 V
19 15 V
20 17 V
19 25 V
20 25 V
20 17 V
19 15 V
LT1
2394 1414 M
180 0 V
750 340 M
19 17 V
20 16 V
19 15 V
20 14 V
19 13 V
20 12 V
20 12 V
19 12 V
20 10 V
19 11 V
20 10 V
19 10 V
20 9 V
20 9 V
19 9 V
20 9 V
19 8 V
20 8 V
19 8 V
20 8 V
20 8 V
19 8 V
20 7 V
19 7 V
20 7 V
19 7 V
20 7 V
20 7 V
19 7 V
20 6 V
19 7 V
20 6 V
19 7 V
20 6 V
20 6 V
19 6 V
20 6 V
19 6 V
20 6 V
19 6 V
20 6 V
20 5 V
19 6 V
20 5 V
19 6 V
20 5 V
19 6 V
20 5 V
20 5 V
19 6 V
20 5 V
19 5 V
20 5 V
19 5 V
20 5 V
20 5 V
19 5 V
20 5 V
19 5 V
20 5 V
19 5 V
20 4 V
20 5 V
19 5 V
20 4 V
19 5 V
20 5 V
19 4 V
20 5 V
20 4 V
19 5 V
20 4 V
19 4 V
20 5 V
19 4 V
20 5 V
20 4 V
19 4 V
20 4 V
19 5 V
20 4 V
19 4 V
20 4 V
20 4 V
19 4 V
20 5 V
19 4 V
20 4 V
19 4 V
20 4 V
20 4 V
19 4 V
20 4 V
19 4 V
20 4 V
19 3 V
20 4 V
20 4 V
19 4 V
LT3
2394 1314 M
180 0 V
750 276 M
19 5 V
20 5 V
19 6 V
20 6 V
19 7 V
20 6 V
20 7 V
19 8 V
20 7 V
19 8 V
20 8 V
19 8 V
20 8 V
20 9 V
19 9 V
20 9 V
19 9 V
20 10 V
19 9 V
20 10 V
20 10 V
19 10 V
20 11 V
19 10 V
20 11 V
19 11 V
20 11 V
20 12 V
19 11 V
20 12 V
19 12 V
20 12 V
19 12 V
20 12 V
20 13 V
19 12 V
20 13 V
19 13 V
20 13 V
19 13 V
20 14 V
20 13 V
19 14 V
20 14 V
19 14 V
20 14 V
19 14 V
20 15 V
20 14 V
19 15 V
20 15 V
19 15 V
20 15 V
19 15 V
20 15 V
20 16 V
19 16 V
20 15 V
19 16 V
20 16 V
19 17 V
20 16 V
20 16 V
19 17 V
20 16 V
19 17 V
20 17 V
19 17 V
20 17 V
20 18 V
19 17 V
20 17 V
19 18 V
20 18 V
19 18 V
20 18 V
20 18 V
19 18 V
20 18 V
19 19 V
20 18 V
19 19 V
20 19 V
20 19 V
19 19 V
20 19 V
19 19 V
20 19 V
19 20 V
20 19 V
20 20 V
19 20 V
20 20 V
19 20 V
20 20 V
19 20 V
20 20 V
20 20 V
19 21 V
750 424 M
19 12 V
20 11 V
19 11 V
20 10 V
19 10 V
20 9 V
20 10 V
19 8 V
20 9 V
19 8 V
20 9 V
19 8 V
20 7 V
20 8 V
19 7 V
20 8 V
19 7 V
20 7 V
19 7 V
20 6 V
20 7 V
19 7 V
20 6 V
19 6 V
20 7 V
19 6 V
20 6 V
20 6 V
19 6 V
20 6 V
19 5 V
20 6 V
19 6 V
20 5 V
20 6 V
19 5 V
20 6 V
19 5 V
20 5 V
19 6 V
20 5 V
20 5 V
19 5 V
20 5 V
19 5 V
20 5 V
19 5 V
20 5 V
20 5 V
19 4 V
20 5 V
19 5 V
20 5 V
19 4 V
20 5 V
20 4 V
19 5 V
20 5 V
19 4 V
20 4 V
19 5 V
20 4 V
20 5 V
19 4 V
20 4 V
19 4 V
20 5 V
19 4 V
20 4 V
20 4 V
19 4 V
20 5 V
19 4 V
20 4 V
19 4 V
20 4 V
20 4 V
19 4 V
20 4 V
19 4 V
20 4 V
19 4 V
20 3 V
20 4 V
19 4 V
20 4 V
19 4 V
20 4 V
19 3 V
20 4 V
20 4 V
19 3 V
20 4 V
19 4 V
20 4 V
19 3 V
20 4 V
20 3 V
19 4 V
stroke
grestore
end
showpage
}
\put(2334,1314){\makebox(0,0)[r]{{\small Asymptots}}}
\put(2334,1414){\makebox(0,0)[r]{{\small Approximation b')}}}
\put(2334,1514){\makebox(0,0)[r]{{\small Unapproximated}}}
\put(1648,51){\makebox(0,0){$\kappa$}}
\put(100,964){%
\special{ps: gsave currentpoint currentpoint translate
270 rotate neg exch neg exch translate}%
\makebox(0,0)[b]{\shortstack{The unintegrated structure function}}%
\special{ps: currentpoint grestore moveto}%
}
\put(2557,151){\makebox(0,0){14}}
\put(2278,151){\makebox(0,0){12}}
\put(1998,151){\makebox(0,0){10}}
\put(1718,151){\makebox(0,0){8}}
\put(1439,151){\makebox(0,0){6}}
\put(1159,151){\makebox(0,0){4}}
\put(880,151){\makebox(0,0){2}}
\put(600,151){\makebox(0,0){0}}
\put(540,1677){\makebox(0,0)[r]{100}}
\put(540,1392){\makebox(0,0)[r]{80}}
\put(540,1107){\makebox(0,0)[r]{60}}
\put(540,821){\makebox(0,0)[r]{40}}
\put(540,536){\makebox(0,0)[r]{20}}
\put(540,251){\makebox(0,0)[r]{0}}
\end{picture}
\caption{The solution $f(\kappa)$ to the integral eqaution 
(\ref{masterf})
(solid line) and for the approximation b') (dashed line) together with
the asymptotic forms.}
\label{asymps}
\end{figure}

In Fig \ref{asymps} we show the solution to the original equation 
(\ref{masterf}) together
with approximation b'), which corresponds to the replacement 
$h \rightarrow 1$.
It is easy to demonstrate that in all cases $f(\kappa)$ and 
${\cal F}$ have asymptotically the same powerlike behaviour 
as the simpler approximation studied in section \ref{quaprop}, Eqs
(\ref{fasymptot}) and (\ref{asymptotcalF}).

We note, however, that the parameter $\lambda$ takes on 
different values in the different approximations. 
If we calculate the first correction term to $f$ in an expansion 
in $1/\kappa $
\be
\label{asymf}
f(\kappa)\propto \kappa^{\frac{\alpha_0}{\lambda}-1}\left( 1+ 
b \kappa^{-1}+ {\cal O}(\kappa^{-2}) \right)
\ee
we obtain
\be
b=-\frac{\alpha_0 (\alpha_0 - \lambda)}{\lambda} \left[ \frac{1}{1+
\lambda}-{\cal A} \right] & \mbox{where} & {\cal A}=
\int(1-h(\Delta))d\Delta \approx 0.65
\ee
For $\lambda \approx 0.4$,
$b$ is very small, and from Fig \ref{asymps} we see that the single power 
in Eq (\ref{fasymptot}) is a surprisingly good approximation 
in the whole $\kappa$-range. 
(For the approximation b') with $h \rightarrow 1$, we 
have ${\cal A} = 0$, and the correction term becomes much larger, 
as is also seen in Fig \ref{asymps}.)

\begin{figure}

\setlength{\unitlength}{0.1bp}
\special{!
/gnudict 40 dict def
gnudict begin
/Color false def
/Solid false def
/gnulinewidth 5.000 def
/vshift -33 def
/dl {10 mul} def
/hpt 31.5 def
/vpt 31.5 def
/M {moveto} bind def
/L {lineto} bind def
/R {rmoveto} bind def
/V {rlineto} bind def
/vpt2 vpt 2 mul def
/hpt2 hpt 2 mul def
/Lshow { currentpoint stroke M
  0 vshift R show } def
/Rshow { currentpoint stroke M
  dup stringwidth pop neg vshift R show } def
/Cshow { currentpoint stroke M
  dup stringwidth pop -2 div vshift R show } def
/DL { Color {setrgbcolor Solid {pop []} if 0 setdash }
 {pop pop pop Solid {pop []} if 0 setdash} ifelse } def
/BL { stroke gnulinewidth 2 mul setlinewidth } def
/AL { stroke gnulinewidth 2 div setlinewidth } def
/PL { stroke gnulinewidth setlinewidth } def
/LTb { BL [] 0 0 0 DL } def
/LTa { AL [1 dl 2 dl] 0 setdash 0 0 0 setrgbcolor } def
/LT0 { PL [] 0 1 0 DL } def
/LT1 { PL [4 dl 2 dl] 0 0 1 DL } def
/LT2 { PL [2 dl 3 dl] 1 0 0 DL } def
/LT3 { PL [1 dl 1.5 dl] 1 0 1 DL } def
/LT4 { PL [5 dl 2 dl 1 dl 2 dl] 0 1 1 DL } def
/LT5 { PL [4 dl 3 dl 1 dl 3 dl] 1 1 0 DL } def
/LT6 { PL [2 dl 2 dl 2 dl 4 dl] 0 0 0 DL } def
/LT7 { PL [2 dl 2 dl 2 dl 2 dl 2 dl 4 dl] 1 0.3 0 DL } def
/LT8 { PL [2 dl 2 dl 2 dl 2 dl 2 dl 2 dl 2 dl 4 dl] 0.5 0.5 0.5 DL } def
/P { stroke [] 0 setdash
  currentlinewidth 2 div sub M
  0 currentlinewidth V stroke } def
/D { stroke [] 0 setdash 2 copy vpt add M
  hpt neg vpt neg V hpt vpt neg V
  hpt vpt V hpt neg vpt V closepath stroke
  P } def
/A { stroke [] 0 setdash vpt sub M 0 vpt2 V
  currentpoint stroke M
  hpt neg vpt neg R hpt2 0 V stroke
  } def
/B { stroke [] 0 setdash 2 copy exch hpt sub exch vpt add M
  0 vpt2 neg V hpt2 0 V 0 vpt2 V
  hpt2 neg 0 V closepath stroke
  P } def
/C { stroke [] 0 setdash exch hpt sub exch vpt add M
  hpt2 vpt2 neg V currentpoint stroke M
  hpt2 neg 0 R hpt2 vpt2 V stroke } def
/T { stroke [] 0 setdash 2 copy vpt 1.12 mul add M
  hpt neg vpt -1.62 mul V
  hpt 2 mul 0 V
  hpt neg vpt 1.62 mul V closepath stroke
  P  } def
/S { 2 copy A C} def
end
}
\begin{picture}(2880,1728)(0,0)
\special{"
gnudict begin
gsave
50 50 translate
0.100 0.100 scale
0 setgray
/Helvetica findfont 100 scalefont setfont
newpath
-500.000000 -500.000000 translate
LTa
LTb
600 251 M
63 0 V
2034 0 R
-63 0 V
600 489 M
63 0 V
2034 0 R
-63 0 V
600 726 M
63 0 V
2034 0 R
-63 0 V
600 964 M
63 0 V
2034 0 R
-63 0 V
600 1202 M
63 0 V
2034 0 R
-63 0 V
600 1439 M
63 0 V
2034 0 R
-63 0 V
600 1677 M
63 0 V
2034 0 R
-63 0 V
600 251 M
0 63 V
0 1363 R
0 -63 V
1019 251 M
0 63 V
0 1363 R
0 -63 V
1439 251 M
0 63 V
0 1363 R
0 -63 V
1858 251 M
0 63 V
0 1363 R
0 -63 V
2278 251 M
0 63 V
0 1363 R
0 -63 V
2697 251 M
0 63 V
0 1363 R
0 -63 V
600 251 M
2097 0 V
0 1426 V
-2097 0 V
600 251 L
LT0
2394 1514 M
180 0 V
819 268 M
19 2 V
19 3 V
19 2 V
19 3 V
19 2 V
18 3 V
19 3 V
19 4 V
19 3 V
19 3 V
19 4 V
19 4 V
18 4 V
19 4 V
19 5 V
19 4 V
19 5 V
19 5 V
19 6 V
19 5 V
18 6 V
19 6 V
19 6 V
19 6 V
19 7 V
19 7 V
19 7 V
19 7 V
18 8 V
19 7 V
19 9 V
19 8 V
19 8 V
19 9 V
19 9 V
19 9 V
18 10 V
19 10 V
19 10 V
19 10 V
19 11 V
19 11 V
19 11 V
19 12 V
18 11 V
19 12 V
19 13 V
19 12 V
19 13 V
19 14 V
19 13 V
19 14 V
18 14 V
19 14 V
19 15 V
19 15 V
19 15 V
19 16 V
19 16 V
19 16 V
18 17 V
19 17 V
19 17 V
19 18 V
19 18 V
19 18 V
19 19 V
19 19 V
18 19 V
19 19 V
19 20 V
19 21 V
19 20 V
19 21 V
19 22 V
18 21 V
19 22 V
19 23 V
19 23 V
19 23 V
19 23 V
19 24 V
19 24 V
18 25 V
19 25 V
19 25 V
19 26 V
19 26 V
19 27 V
19 26 V
19 28 V
18 27 V
19 28 V
19 29 V
19 28 V
19 30 V
19 29 V
19 30 V
19 30 V
LT1
2394 1414 M
180 0 V
819 252 M
19 1 V
19 1 V
19 2 V
19 1 V
19 2 V
18 2 V
19 1 V
19 2 V
19 2 V
19 2 V
19 3 V
19 2 V
18 3 V
19 2 V
19 3 V
19 3 V
19 4 V
19 3 V
19 4 V
19 3 V
18 4 V
19 5 V
19 4 V
19 5 V
19 5 V
19 5 V
19 5 V
19 5 V
18 6 V
19 6 V
19 6 V
19 6 V
19 7 V
19 7 V
19 7 V
19 7 V
18 8 V
19 7 V
19 8 V
19 9 V
19 8 V
19 9 V
19 9 V
19 9 V
18 10 V
19 10 V
19 10 V
19 10 V
19 11 V
19 11 V
19 11 V
19 11 V
18 12 V
19 12 V
19 12 V
19 13 V
19 13 V
19 13 V
19 14 V
19 14 V
18 14 V
19 14 V
19 15 V
19 15 V
19 15 V
19 16 V
19 16 V
19 16 V
18 17 V
19 17 V
19 17 V
19 18 V
19 18 V
19 18 V
19 18 V
18 19 V
19 20 V
19 19 V
19 20 V
19 21 V
19 20 V
19 21 V
19 22 V
18 22 V
19 22 V
19 22 V
19 23 V
19 23 V
19 23 V
19 24 V
19 24 V
18 25 V
19 25 V
19 25 V
19 26 V
19 26 V
19 27 V
19 27 V
19 27 V
LT3
2394 1314 M
180 0 V
819 262 M
19 2 V
19 2 V
19 2 V
19 2 V
19 3 V
18 2 V
19 3 V
19 3 V
19 3 V
19 4 V
19 3 V
19 4 V
18 4 V
19 4 V
19 5 V
19 4 V
19 5 V
19 5 V
19 5 V
19 6 V
18 6 V
19 6 V
19 6 V
19 6 V
19 7 V
19 7 V
19 7 V
19 7 V
18 8 V
19 7 V
19 8 V
19 9 V
19 8 V
19 9 V
19 9 V
19 10 V
18 9 V
19 10 V
19 10 V
19 11 V
19 11 V
19 11 V
19 11 V
19 11 V
18 12 V
19 12 V
19 13 V
19 13 V
19 13 V
19 13 V
19 14 V
19 14 V
18 14 V
19 14 V
19 15 V
19 15 V
19 16 V
19 16 V
19 16 V
19 16 V
18 17 V
19 17 V
19 18 V
19 17 V
19 18 V
19 19 V
19 18 V
19 20 V
18 19 V
19 20 V
19 20 V
19 20 V
19 21 V
19 21 V
19 22 V
18 22 V
19 22 V
19 23 V
19 22 V
19 24 V
19 23 V
19 24 V
19 25 V
18 25 V
19 25 V
19 25 V
19 26 V
19 26 V
19 27 V
19 27 V
19 28 V
18 27 V
19 29 V
19 28 V
19 29 V
19 30 V
19 29 V
19 30 V
19 31 V
stroke
grestore
end
showpage
}
\put(2334,1314){\makebox(0,0)[r]{{The asymptotic form in Eq (\ref{Fexpa})}}}
\put(2334,1414){\makebox(0,0)[r]{{The first term in Eq (\ref{factF})}}}
\put(2334,1514){\makebox(0,0)[r]{{The structure function}}}
\put(1648,51){\makebox(0,0){$\ln Q^2$}}
\put(100,964){%
\special{ps: gsave currentpoint currentpoint translate
270 rotate neg exch neg exch translate}%
\makebox(0,0)[b]{\shortstack{The structure function}}%
\special{ps: currentpoint grestore moveto}%
}
\put(2697,151){\makebox(0,0){10}}
\put(2278,151){\makebox(0,0){8}}
\put(1858,151){\makebox(0,0){6}}
\put(1439,151){\makebox(0,0){4}}
\put(1019,151){\makebox(0,0){2}}
\put(600,151){\makebox(0,0){0}}
\put(540,1677){\makebox(0,0)[r]{120}}
\put(540,1439){\makebox(0,0)[r]{100}}
\put(540,1202){\makebox(0,0)[r]{80}}
\put(540,964){\makebox(0,0)[r]{60}}
\put(540,726){\makebox(0,0)[r]{40}}
\put(540,489){\makebox(0,0)[r]{20}}
\put(540,251){\makebox(0,0)[r]{0}}
\end{picture}
\caption{The $\ln Q^2$-dependence of the structure function. The dashed line
is the contribution from the first term in Eq. (\ref{factF}). The
dotted line is the contribution from the two first terms in the power
expansion in Eq (\ref{Fexpa}).}
\label{F}
\end{figure}
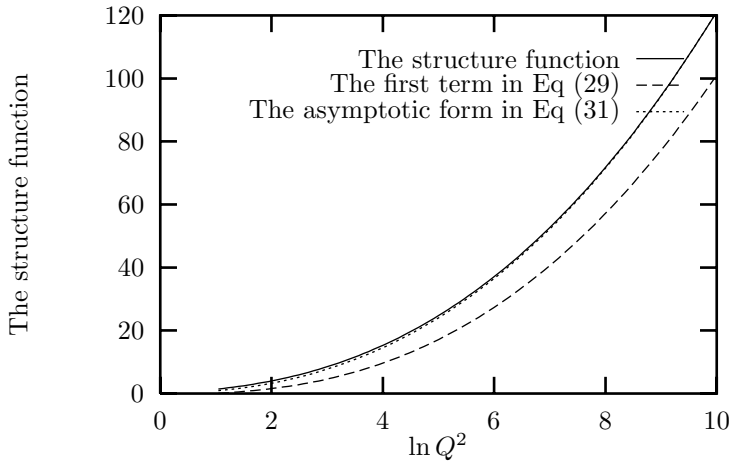

We note that the powerlike dependence on $\kappa$ for 
$f$ and ${\cal F}$ implies a similar power dependence also 
for the original ``integrated'' structure function $F$. 
For small $x$ we have from Eqs (\ref{factform}) and (\ref{defcalF}) 

\begin{eqnarray}
\label{factF}
F&=&x^{-\lambda} \tilde{F}(\ln Q^2) \nonumber \\
\tilde{F} = \int_{\kappa_{cut}}^{\ln(Q^2)} f(\kappa) d\kappa &+& \int_{\ln(Q^2)} e^{(1+\lambda)(\ln(Q^2) - \kappa)} f(\kappa) d\kappa
\end{eqnarray}
The result is presented in Fig \ref{F}, which also shows the separate contributions in Eq (\ref{factF}). These contributions correspond to chains with $k_{\perp}^2 < Q^2$ and $k_{\perp}^2 > Q^2$ respectively, where $k_{\perp}$ is the transverse momentum of the last link in the chain.

Inserting the asymptotic expression in Eq (\ref{asymf}) we find for large $Q^2$  

\be
\label{asymF}
F \propto x^{-\lambda} (\ln Q^2)^{\alpha_0 / \lambda}; \,\,\,\,\ln(1/x) \gg \ln(Q^2) \gg 1
\ee
For large $Q^2$ the first integral in eq (\ref{factF}) dominates, and for this term the single power in Eq (\ref{asymF}) is a good approximation. The second integral is suppressed by the factor $\frac{\alpha_0}{\lambda(1+\lambda)} \cdot \frac{1}{\kappa}$, which for $\ln Q^2 = 10$ is about 20\%
, but becomes relatively more important for smaller $Q^2$. Thus, if we use the power expansion of $f$ in Eq (\ref{asymf}), we get for $b\approx 0$ the better estimate
\begin{equation}
\label{Fexpa}
\tilde{F}\propto (\ln Q^2)^\frac{\alpha_0}{\lambda}
\left[
1+\frac{\alpha_0}{\lambda(1+\lambda)} \cdot \frac{1}{\ln Q^2}
+{\cal O}\left((\ln Q^2)^{-2}\right)
\right]
\end{equation}

To better illustrate the $k_\perp$-dependence of the chain, 
we have studied
a very long chain in order to find out where this chain passes 
the central line
$y=0$. It is straight forward to show that the density $\rho$
(in the (y,$\kappa$)-plane) of all the right hand endpoints of 
the horizontal
lines in the diagram in Fig \ref{triangularfandiagram}, is given by the expression
\begin{equation}
\label{rhodefine}
\rho = F(\ell_1,\kappa)\cdot {\cal F}(\ell_2,\kappa)\cdot \exp(-\kappa)
\end{equation}
where $\ell_1=\ln W -\kappa/2+y$, $\ell_2=\ln W-\kappa/2-y$ and 
$W$ equals
the total cms energy of the chain. The non-integrated function 
${\cal F}(\ell_2,\kappa)$ describes the weight for the chains 
coming from left in Fig \ref{triangularfandiagram} to the point $(y,\kappa)$, 
while the normal structure function $F(\ell_1,\kappa)$ 
adds up the contributions from links with all possible values 
of $\kappa^{\prime}$, both those larger than $\kappa$ and 
those smaller than $\kappa$. Inserting the asymptotic 
expressions in Eqs (\ref{asymptotcalF}) and (\ref{asymF}) 
we find the following result (the factor $\exp (-\lambda \kappa)$
comes from the $\kappa$-dependences of $\ell_1$ and $\ell_2$)

\be
\label{rhoasym}
\rho \sim \kappa^{\frac{2\alpha_0}{\lambda}-1} 
e^{-(1+\lambda)\kappa};\,\,\,W \,  \mbox{and}  \,\kappa \,\,\, 
\mbox{large} 
\ee

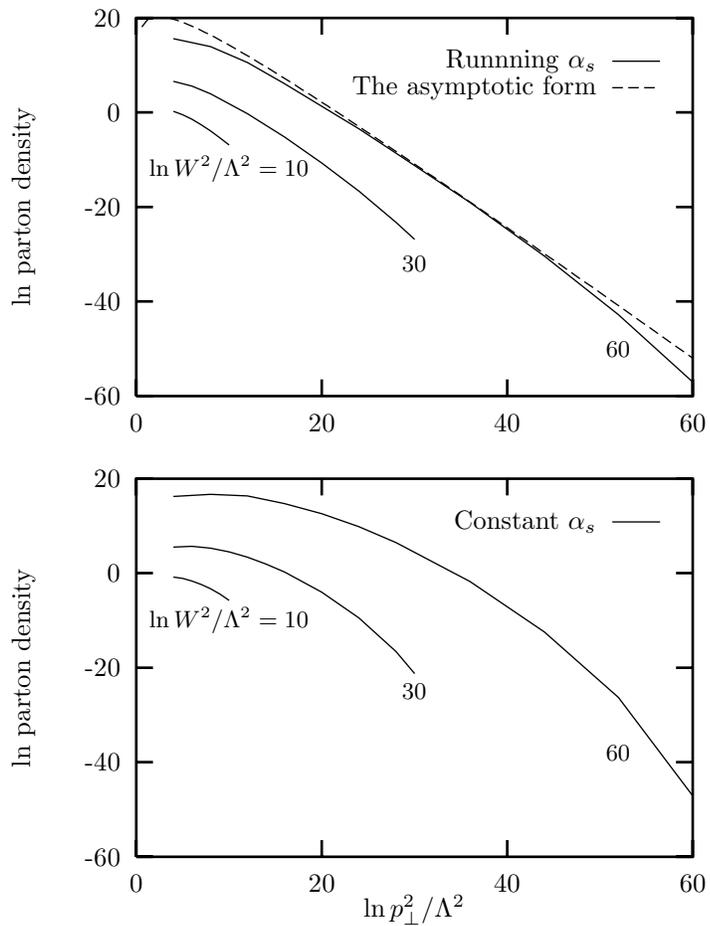
\begin{figure}
\setlength{\unitlength}{0.1bp}
\special{!
/gnudict 40 dict def
gnudict begin
/Color false def
/Solid false def
/gnulinewidth 5.000 def
/vshift -33 def
/dl {10 mul} def
/hpt 31.5 def
/vpt 31.5 def
/M {moveto} bind def
/L {lineto} bind def
/R {rmoveto} bind def
/V {rlineto} bind def
/vpt2 vpt 2 mul def
/hpt2 hpt 2 mul def
/Lshow { currentpoint stroke M
  0 vshift R show } def
/Rshow { currentpoint stroke M
  dup stringwidth pop neg vshift R show } def
/Cshow { currentpoint stroke M
  dup stringwidth pop -2 div vshift R show } def
/DL { Color {setrgbcolor Solid {pop []} if 0 setdash }
 {pop pop pop Solid {pop []} if 0 setdash} ifelse } def
/BL { stroke gnulinewidth 2 mul setlinewidth } def
/AL { stroke gnulinewidth 2 div setlinewidth } def
/PL { stroke gnulinewidth setlinewidth } def
/LTb { BL [] 0 0 0 DL } def
/LTa { AL [1 dl 2 dl] 0 setdash 0 0 0 setrgbcolor } def
/LT0 { PL [] 0 1 0 DL } def
/LT1 { PL [4 dl 2 dl] 0 0 1 DL } def
/LT2 { PL [2 dl 3 dl] 1 0 0 DL } def
/LT3 { PL [1 dl 1.5 dl] 1 0 1 DL } def
/LT4 { PL [5 dl 2 dl 1 dl 2 dl] 0 1 1 DL } def
/LT5 { PL [4 dl 3 dl 1 dl 3 dl] 1 1 0 DL } def
/LT6 { PL [2 dl 2 dl 2 dl 4 dl] 0 0 0 DL } def
/LT7 { PL [2 dl 2 dl 2 dl 2 dl 2 dl 4 dl] 1 0.3 0 DL } def
/LT8 { PL [2 dl 2 dl 2 dl 2 dl 2 dl 2 dl 2 dl 4 dl] 0.5 0.5 0.5 DL } def
/P { stroke [] 0 setdash
  currentlinewidth 2 div sub M
  0 currentlinewidth V stroke } def
/D { stroke [] 0 setdash 2 copy vpt add M
  hpt neg vpt neg V hpt vpt neg V
  hpt vpt V hpt neg vpt V closepath stroke
  P } def
/A { stroke [] 0 setdash vpt sub M 0 vpt2 V
  currentpoint stroke M
  hpt neg vpt neg R hpt2 0 V stroke
  } def
/B { stroke [] 0 setdash 2 copy exch hpt sub exch vpt add M
  0 vpt2 neg V hpt2 0 V 0 vpt2 V
  hpt2 neg 0 V closepath stroke
  P } def
/C { stroke [] 0 setdash exch hpt sub exch vpt add M
  hpt2 vpt2 neg V currentpoint stroke M
  hpt2 neg 0 R hpt2 vpt2 V stroke } def
/T { stroke [] 0 setdash 2 copy vpt 1.12 mul add M
  hpt neg vpt -1.62 mul V
  hpt 2 mul 0 V
  hpt neg vpt 1.62 mul V closepath stroke
  P  } def
/S { 2 copy A C} def
end
}
\begin{picture}(2880,1728)(0,0)
\special{"
gnudict begin
gsave
50 50 translate
0.100 0.100 scale
0 setgray
/Helvetica findfont 100 scalefont setfont
newpath
-500.000000 -500.000000 translate
LTa
LTb
600 251 M
63 0 V
2034 0 R
-63 0 V
600 608 M
63 0 V
2034 0 R
-63 0 V
600 964 M
63 0 V
2034 0 R
-63 0 V
600 1321 M
63 0 V
2034 0 R
-63 0 V
600 1677 M
63 0 V
2034 0 R
-63 0 V
600 251 M
0 63 V
0 1363 R
0 -63 V
1299 251 M
0 63 V
0 1363 R
0 -63 V
1998 251 M
0 63 V
0 1363 R
0 -63 V
2697 251 M
0 63 V
0 1363 R
0 -63 V
600 251 M
2097 0 V
0 1426 V
-2097 0 V
600 251 L
LT0
2394 1514 M
180 0 V
740 1325 M
35 -13 V
35 -17 V
35 -21 V
35 -23 V
35 -26 V
35 -27 V
740 1438 M
70 -18 V
70 -29 V
139 -75 V
140 -88 V
140 -98 V
140 -106 V
1579 906 L
70 -64 V
740 1599 M
140 -30 V
139 -60 V
140 -80 V
140 -85 V
140 -85 V
140 -92 V
1858 981 L
2138 779 L
2417 558 L
2697 304 L
LT1
2394 1414 M
180 0 V
621 1643 M
21 27 V
15 7 V
65 0 R
5 -1 V
21 -6 V
21 -7 V
22 -9 V
21 -9 V
21 -9 V
21 -11 V
21 -10 V
22 -11 V
21 -12 V
21 -11 V
21 -12 V
21 -12 V
21 -13 V
22 -12 V
21 -13 V
21 -13 V
21 -12 V
21 -13 V
22 -14 V
21 -13 V
21 -13 V
21 -13 V
21 -14 V
21 -14 V
22 -13 V
21 -14 V
21 -14 V
21 -13 V
21 -14 V
22 -14 V
21 -14 V
21 -14 V
21 -14 V
21 -14 V
21 -14 V
22 -14 V
21 -15 V
21 -14 V
21 -14 V
21 -14 V
22 -15 V
21 -14 V
21 -15 V
21 -14 V
21 -14 V
21 -15 V
22 -14 V
21 -15 V
21 -14 V
21 -15 V
21 -14 V
22 -15 V
21 -15 V
21 -14 V
21 -15 V
21 -14 V
21 -15 V
22 -15 V
21 -15 V
21 -14 V
21 -15 V
21 -15 V
22 -14 V
21 -15 V
21 -15 V
21 -15 V
21 -15 V
21 -14 V
22 -15 V
21 -15 V
21 -15 V
21 -15 V
21 -15 V
22 -15 V
21 -14 V
21 -15 V
21 -15 V
21 -15 V
21 -15 V
22 -15 V
21 -15 V
21 -15 V
21 -15 V
21 -15 V
22 -15 V
21 -15 V
21 -15 V
21 -15 V
21 -15 V
21 -15 V
22 -15 V
21 -15 V
21 -15 V
stroke
grestore
end
showpage
}
\put(2334,1414){\makebox(0,0)[r]{The asymptotic form}}
\put(2334,1514){\makebox(0,0)[r]{{Runnning $\alpha_s$}}}
\put(2417,429){\makebox(0,0){{\small 60}}}
\put(1649,750){\makebox(0,0){{\small 30}}}
\put(950,1107){\makebox(0,0){{\small $\ln W^2/\Lambda^2 = 10$}}}
\put(1648,51){\makebox(0,0){\\}}
\put(220,964){%
\special{ps: gsave currentpoint currentpoint translate
270 rotate neg exch neg exch translate}%
\makebox(0,0)[b]{\shortstack{ln parton density}}%
\special{ps: currentpoint grestore moveto}%
}
\put(2697,151){\makebox(0,0){60}}
\put(1998,151){\makebox(0,0){40}}
\put(1299,151){\makebox(0,0){20}}
\put(600,151){\makebox(0,0){0}}
\put(540,1677){\makebox(0,0)[r]{20}}
\put(540,1321){\makebox(0,0)[r]{0}}
\put(540,964){\makebox(0,0)[r]{-20}}
\put(540,608){\makebox(0,0)[r]{-40}}
\put(540,251){\makebox(0,0)[r]{-60}}
\end{picture}
\begin{picture}(2880,1728)(0,0)
\special{"
gnudict begin
gsave
50 50 translate
0.100 0.100 scale
0 setgray
/Helvetica findfont 100 scalefont setfont
newpath
-500.000000 -500.000000 translate
LTa
LTb
600 251 M
63 0 V
2034 0 R
-63 0 V
600 608 M
63 0 V
2034 0 R
-63 0 V
600 964 M
63 0 V
2034 0 R
-63 0 V
600 1321 M
63 0 V
2034 0 R
-63 0 V
600 1677 M
63 0 V
2034 0 R
-63 0 V
600 251 M
0 63 V
0 1363 R
0 -63 V
1299 251 M
0 63 V
0 1363 R
0 -63 V
1998 251 M
0 63 V
0 1363 R
0 -63 V
2697 251 M
0 63 V
0 1363 R
0 -63 V
600 251 M
2097 0 V
0 1426 V
-2097 0 V
600 251 L
LT0
2394 1514 M
180 0 V
740 1306 M
35 -5 V
35 -10 V
35 -13 V
35 -16 V
35 -20 V
35 -24 V
740 1419 M
70 3 V
70 -7 V
70 -14 V
69 -20 V
70 -26 V
70 -30 V
140 -76 V
140 -97 V
140 -127 V
70 -83 V
740 1610 M
140 8 V
139 -6 V
140 -29 V
140 -38 V
140 -49 V
140 -60 V
279 -147 V
280 -190 V
2417 852 L
2697 481 L
stroke
grestore
end
showpage
}
\put(2334,1514){\makebox(0,0)[r]{{Constant $\alpha_s$}}}
\put(2417,643){\makebox(0,0){{\small 60}}}
\put(1649,875){\makebox(0,0){{\small 30}}}
\put(950,1142){\makebox(0,0){{\small $\ln W^2/\Lambda^2 = 10$}}}
\put(1648,51){\makebox(0,0){$\ln p_\perp^2/\Lambda^2$}}
\put(220,964){%
\special{ps: gsave currentpoint currentpoint translate
270 rotate neg exch neg exch translate}%
\makebox(0,0)[b]{\shortstack{ln parton density}}%
\special{ps: currentpoint grestore moveto}%
}
\put(2697,151){\makebox(0,0){60}}
\put(1998,151){\makebox(0,0){40}}
\put(1299,151){\makebox(0,0){20}}
\put(600,151){\makebox(0,0){0}}
\put(540,1677){\makebox(0,0)[r]{20}}
\put(540,1321){\makebox(0,0)[r]{0}}
\put(540,964){\makebox(0,0)[r]{-20}}
\put(540,608){\makebox(0,0)[r]{-40}}
\put(540,251){\makebox(0,0)[r]{-60}}
\end{picture}
\caption{Transverse momentum distributions in the centre of a 
long chain for different values of hadronic central mass $W$ for running vs. fixed 
$\alpha_s$.}
\label{rho_kappa}
\end{figure}

In Fig \ref{rho_kappa} we present the results from a 
Monte Carlo simulation for different energies $W$ of the quantity
$\rho$ in Eq (\ref{rhodefine}). We note that the distribution 
has the expected approximate exponential fall off 
$\sim \exp[-(1+\lambda)\kappa]$, in accordance with Eq (\ref{rhoasym})
and this corresponds to a (negative)
power dependence
in $k_{\perp}$ for the density $\rho$. Further the curve has the
same slope independent of the total energy $W$, 
consistent with an
asymptotically factorizing dependence on $x$ and $k_\perp$, i.e. there
is no sign of the expected BFKL diffusion in $\kappa$. 
The dashed line in Fig \ref{rho_kappa} shows the asymptotic form in 
Eq (\ref{rhoasym}).

For comparison we also show in Fig \ref{rho_kappa} the result 
for a constant
coupling ${\bar \alpha}=0.2$. Here we find a Gaussian 
distribution with
a width increasing with the energy ($\sim \sqrt{\ln 1/x}$), 
as expected from a random walk.

\section{Dominating Paths in Transverse Momentum}
\label{dominpath}

The result in Eq (\ref{asymF}) is particularly interesting if we
compare to a result obtained in \cite{JSHKBAGG}, where 
we study the mean paths of the initial state cascade
in the $(\ell,\kappa)$-plane. We find that due to the running coupling
the region covered by the major contributions (nowadays known as the
``Bartel cigar'') corresponds to one part, length $\ell_1$, 
with small $\kappa$-values
and with a BFKL-like contribution, $\simeq$$\exp(\lambda \ell_1)$ and 
a second part,
length $\ell_2=\ell-\ell_1$, corresponding to increasing
$\kappa$-values up to the final $\kappa_Q\equiv \ln(Q^2)$. This
second part is a DGLAP
contribution, $\simeq$$\exp(2\sqrt{\alpha_0\ell_2 \chi_Q})$, where
$\chi_Q=\ln(\kappa_Q)$.
 Using a saddle-point 
approximation we demonstrated that the combination of the 
two parts corresponds for $\lambda \ell>\alpha_0 \chi_Q/ \lambda$ 
to 
\be
{\cal F}\sim\exp(\lambda\ell+\alpha_0\chi_Q/\lambda)
\ee
which is identical to our result from Eq (\ref{asymF}).
 
For  $\lambda \ell<\alpha_0\chi_Q/\lambda$ the DGLAP chains dominate, 
because in this case the saddle-point is not within the the 
allowed integration region. We note that this result 
implies that the BFKL behaviour is relevant only for 
rather small $x$-values, $x\sim10^{-5}$ for $Q^2\sim 100\ {\rm GeV}^2$.

We will end by showing that for small
$x$-values the very construction of the
LDC chains leads to a simple result of the BFKL kind. 
 In the LDC model it is 
 possible to go
``up'' and ``down'' in $\kappa$  along the chains, and thus the total chain can be divided into "cells", 
where each cell contains a set of up-steps followed by another set of
down-steps. Let us study one such cell, which starts in the point at $(y_1 , \kappa_1)$, goes up to a maximum point $(y^{\prime}, \kappa^{\prime})$, and then goes down to the endpoint $(y_2 , \kappa_2)$. We note that this 
second part
corresponds to up-steps from the probe side, and therefore both pieces correspond to "DGLAP-chains".

To describe the path we generalize the structure function ${\cal F}$ to a function ${\cal F}(\ell, \kappa_2, \kappa_1)$, which describes the sum of all paths, which start at a virtuality given by $\kappa_1$ (which may now be different from $\kappa_{cut}$) and end at $\kappa_2$. It is also convenient to include the exponential factor in Eq (\ref{rhodefine}) into the structure function, and define a "symmetrized structure function" ${\cal F}_s$  by the relation

\be
{\cal F}_s (\ell, \kappa_2, \kappa_1) = e^{-(\kappa_2 -\kappa_1)/2} {\cal F}
\ee
If expressed in the rapidity separation between the two endpoints, $\delta y = \ell +(\kappa_2 - \kappa_1)/2$, this function is fully symmetric with respect to the direction of the chain. The result for a long chain is also the direct product of the separate pieces.

 Inserting the DGLAP-result for the two sections in the cell, and integrating over possible intermediate points $(y^{\prime},\kappa^{\prime})$ we find

\be
\label{cell}
&&{\cal F}_{s,cell}(\delta y=y_2 - y_1,\kappa_2,\kappa_1) \nonumber \\ &&\approx e^{-(\kappa_2 -\kappa_1)/2} \int d\kappa^{\prime} dy^{\prime} e^{-(\kappa^{\prime}-\kappa)} {\cal F}_{DGLAP}(\ell_1,\kappa^{\prime},\kappa_1) {\cal F}_{DGLAP}(\ell_2,\kappa^{\prime},\kappa_2) \nonumber \\ &&{\mathrm where}\,\,\,
\ell_1 = y^{\prime} -y_1 - (\kappa^{\prime} -\kappa_1)/2;\,\,\, \ell_2 = y - y^{\prime} - (\kappa^{\prime} -\kappa_2)/2
\ee

We start by considering the simpler case with a fixed coupling $\bar{\alpha}$, for which we have

\be
{\cal F}_{DGLAP}(\ell, \kappa^{\prime}, \kappa_1) \simeq
exp(2\sqrt{\bar{\alpha}\ell (\kappa^{\prime}-\kappa_1)})
\ee
Inserted into Eq (\ref{cell}) this gives
\be
\label{hownicethesteps}
{\cal F}_{s,cell} \approx \int dy^{\prime} d\kappa^{\prime}
\exp[2\sqrt{\bar{\alpha}\ell_1(\kappa^{\prime}-\kappa_1)}+ 
2\sqrt{\bar{\alpha} \ell_2 (\kappa^{\prime}-\kappa_2)}-\kappa^{\prime}+
\kappa_2/2+\kappa_1/2]
\ee
where $\ell_1$ and $\ell_2$ are given by Eq (\ref{cell}).
This
integral can be solved by stationary-phase methods, 
i.e. we look for the saddle-point in the integrand exponent. 
After a little algebra we find the surprisingly simple result
that
\be
\label{resulthownice}
{\cal F}_{s,cell} \simeq \exp[(R-1)\delta y/2] & \mbox{with} &
R=\sqrt{1+8\bar{\alpha}} \,\,\,\mbox{and}\,\,\, \delta y = y_2-y_1
\ee
Thus the contribution from one cell depends only upon the
rapidity difference between the starting point and the endpoint.
This also means that 
this result
can be easily generalized to any number of cells.

 A
closer examination tells us, however, that the maximum is only obtained 
within the
integration region if $\delta y$ is large compared to $\delta \kappa= \kappa_2 - \kappa_1$. 
There will be a dividing line with
\be
\label{BFKLsoftline}
\delta \kappa = \frac{R-1}{R} \delta y
\ee
(where $R$ is defined in Eq (\ref{resulthownice})) with 
the property that for
smaller $\kappa$-values there is a saddle-point but for larger 
$\kappa$ the main
contribution is a single DGLAP-motion always directed upwards in $\kappa$. 
 For $\kappa$
below the line the result is obviously of the BFKL kind, i.e.
there is (besides the symmetrical $\kappa$-dependence) an effective
$x^{-\lambda_e}$ behaviour, but this time with 
$\lambda_e=(R-1)/2 \simeq 0.3$ 
for our ``conventional'' value of $\bar{\alpha}\simeq 3/\pi\times 0.2$.

For a running coupling, cells which go very high up in $\kappa$ are suppressed. Therefore the cells will have limited sizes, both in $\kappa$ and in $y$, and for very small $x$ the number of cells will be proportional to the total rapidity range, $\ln (1/x)$. The result becomes a BFKL-like power-dependence on $1/x$ but a saturating distribution in $\kappa$, which is described by the limited $\kappa$-distribution in a single cell, a behaviour which is in agreement with our results in section \ref{running}. The dividing line in Eq (\ref{BFKLsoftline}) will be relevant in the region not too far from the target end, before the $\kappa$-distribution saturates for very large rapidity separations.

\section{Conclusions}

We have shown that if we introduce a running coupling
$\bar{\alpha}=\alpha_0/\kappa$ into the integral equations for the gluon
structure function in the LDC model then

\begin{itemize}

\item for the non-integrated structure function ${\cal F}(\ell\equiv
\ln(1/x),\kappa\equiv \ln(k_{\perp}^2))$ we obtain a factorizing 
BFKL-like behaviour
\be
{\cal F}(\ell,\kappa)\simeq \exp(\lambda_m \ell) f_{\lambda_m}(\kappa)
\ee
where the largest eigenvalue $\lambda_m$ is isolated. It depends upon the
soft cutoff $\kappa_c$ but takes on stable values $\lambda_m\simeq
0.4\, \alpha_0$ for $0.5 \leq \kappa_c\leq 2$. Further a very good
approximation for $f_{\lambda_m}$ valid for $\kappa > \kappa_c$ is
\be
f_{\lambda_m} \simeq \kappa^{\alpha_0/\lambda_m-1}
\ee

\item For the (integrated) structure function $F(\ell\equiv \ln(1/x),
\kappa_Q\equiv \ln(Q^2))$ we then obtain, again as a
good approximation for large $\ell$, ($\lambda_m \ell > \alpha_0
\ln(\kappa_Q)/\lambda_m$)
\be
\label{transition}
F(\ell, \kappa_Q) \sim const\cdot\exp(\lambda_m \ell + \alpha_0
\ln(\kappa_Q)/\lambda_m)
\ee
which, as it is demonstrated in \cite{JSHKBAGG}, corresponds to an
interpolation between the BFKL and DGLAP mechanisms. For $\alpha_0
\ln(\kappa_Q)/\lambda_m >\lambda_m \ell$ the DGLAP mechanism dominates
and there is evidently according to Eq (\ref{transition}) a smooth
turnover to DGLAP when $\alpha_0
\ln(\kappa_Q)/\lambda_m =\lambda_m \ell$.

\item The expected BFKL-diffusion in $\kappa$ (with $\ell$ as the
``time''-variable) is not noticeable in the model. For large values of
$\ell$ we obtain in the center of phase space a saturating
$\kappa$-distribution (valid at least for $\kappa >2$)
\be
\rho d\kappa \simeq d\kappa \kappa^{2 \alpha_0/\lambda_m-1}
\exp\left[-(\lambda_m+1)\kappa\right] 
\simeq \frac{dk_{\perp}^2}{k_{\perp}^{4+ 2 \lambda_m}} 
\bar{\alpha}\cdot (\ln(k_{\perp}^2))^{2\alpha_0/\lambda_m}
\ee
which may be characterized as a kind of ``renormalized Rutherford cross
section''. 

\end{itemize}

\appendix

\section{Numerical Method for Solving the Integral Equation}
\label{numapp}
We will here briefly describe the method of using Chebyshev polynomials
for solving the integral equation (\ref{masterf}).

The Chebyshev  polynomials are defined by

$$
T_n(x)=\cos n(\arccos x), -1<x<1
$$
and obey the orthogonality relation

\begin{equation}
\int_{-1}^1 \frac{T_m(x)T_n(x)}{\sqrt{1-x^2}}=
\left\{
\begin{array}[c]{c}
0, m\neq n\\
\frac{\pi}{2}, m=n\neq 0\\
\pi, m=n=0\\
\end{array}
\right .
\end{equation}

The eigenfunction and the kernel of the integral equation (times a factor
$\sqrt{1-x'^2}$, because of the weight factor in the orthogonality relation)
are expanded
in Chebyshev polynomials, up to a certain degree $N$, thus converting the
integral equation to an approximately equivalent matrix equation. Usually,
we are interested in the largest eigenvalue and the corresponding 
eigenfunction. 

For practical reasons, we must approximate the upper
limit in the integral equation with a large number which
we have set to $\kappa_{max}=100$. No significant
dependence on the value of $\kappa_{max}>50$ has been noted.

Normally, the error in the largest eigenvalue $\lambda$ is less than
1\% for the number of polynomials $N>50$ but 
we have used up to 320 polynomials to get a good description of the
eigenfunctions for large $\kappa$-values. 

In this work,
we have used computer code from Numerical Recipes \cite{NUMREC}.

\section{Analytic Solution to the Approximated Integral Equation}
\label{solution}
In this appendix we will present an analytic solution 
to the integral equation:

\begin{equation}
\lambda f(\kappa )=
\bar\alpha \int_{\kappa_{min}}^\kappa  f(\kappa ') d\kappa'
+e^{(\lambda +1)\kappa} \int_\kappa^{\kappa_{max}}
\bar\alpha f(\kappa ') e^{ -(\lambda +1)\kappa'} d\kappa',
\label{appint}
\end{equation}
with a running coupling $\bar \alpha (\kappa)=\alpha_0/\kappa$. This
corresponds to approximation b') in section \ref{xdep} where $h$
is replaced by 1.

For finite $\kappa_{min}$, $\kappa_{max}$ the integral equation is
solved by rewriting it to a standard differential equation with solutions
which are expressed in confluent hyper geometric functions. Using the
boundary condition at $\kappa_{max}$ and taking the
limit $\kappa_{max} \rightarrow \infty$, we find that solutions to the
integral equation have the asymptotic ($\kappa \rightarrow \infty$) behaviour 
$f(\kappa)\sim \kappa^{\frac{\alpha_0}{\lambda}-1}$.
Using also the boundary condition at $\kappa_{min}$, the largest eigenvalue 
can be found as a function of $\kappa_{min}$. No simple relation for 
$\lambda (\kappa_{min})$ is found so it is evaluated numerically.
The results are in agreement with the numerical solution of the integral
equation.

We start by changing variables to $u=(\lambda+1)\kappa$ to get rid of
$\lambda$ from the right hand side of the integral equation (\ref{appint})
which becomes

\begin{equation}
\frac{\lambda}{\alpha_0} \tilde f(u)=
\frac{1}{u} \int_{u_{min}}^u
\tilde f(u') du'
+
e^u
\int_u^{u_{max}}
\frac{e^{-u'}}{u'} \tilde f(u') du'.
\end{equation}
This can be rewritten to the differential equation

\begin{equation}
\left(
\frac{\tilde f'-\tilde f}{-\frac{1}{u^2}-\frac{1}{u}}
\right)'
=\frac{\alpha_0}{\lambda}\tilde f.
\label{diffop}
\end{equation}
With 
$$
y(u)\equiv\frac{\tilde f'-\tilde f}{-\frac{1}{u^2}-\frac{1}{u}} 
\Rightarrow
f(\kappa)=\tilde f(u)=\frac{\lambda}{\alpha_0}y'(u),
$$
it can be rewritten to the 
following differential equation for $y(u)$

\begin{equation}
y''-y'+\left(\frac{1}{u^2}+\frac{1}{u}\right)\frac{\alpha_0}{\lambda}y=0.
\end{equation}

Two linearly independent solutions are given by the real and imaginary part
of the function (see e.g. Ref \cite{maths})
$$
\tilde M(u)\equiv
\frac{\Gamma\left(\frac{1}{4}-\nu^2+i\nu\right)}
{\Gamma\left(1+i2\nu\right)}
u^{\frac{1}{2}+i\nu}
M\left(\frac{1}{4}-\nu^2+i\nu,1+i2\nu;u\right),
$$
where $\nu \equiv \sqrt{\frac{\alpha_0}{\lambda}-\frac{1}{4}}$ and
$M$ is the standard notation of the confluent hyper geometric function:

$$
M(a,b;u)=\sum \frac{a(a+1)\cdots (a+n-1)}{b(b+1)\cdots (b+n-1)}
\frac{u^n}{n!}.
$$
The real and imaginary parts of $\tilde M$ have the asymptotic
behaviour

\begin{eqnarray}
\label{solutions}
y_1(u)\equiv{\rm Re}\left[\tilde M(u)\right]&\sim& 
u^{-\frac{\alpha_0}{\lambda}}e^u
\nonumber \\
y_2(u)\equiv{\rm Im}\left[\tilde M(u)\right]&\sim& 
u^{\frac{\alpha_0}{\lambda}}
\end{eqnarray}

To find the boundary conditions from the integral
equation we insert
the differential operator in the left hand side of eq. (\ref{diffop}) into 
the integral equation. The boundary conditions are

\begin{eqnarray}
\left[
\frac{\tilde f' -\tilde f}{-\frac{1}{u^2}-\frac{1}{u}}
\right]_{u=u_{min}}&=&0
\nonumber
\\
\left[
e^{-u}\frac{u\tilde f'+\tilde f}{1+u}
\right]_{u=u_{max}}&=&0.
\label{bound}
\end{eqnarray} 
From the second boundary condition, we see that the power solution 
($y_2$ in Eqs. (\ref{solutions}))
becomes more and more important as $u_{max}\rightarrow \infty$.
This means that solutions to the integral equation have the
 asymptotic behaviour 
$f(\kappa)\sim \kappa^{\frac{\alpha_0}{\lambda}-1}$.

Having singled out $y_2$, we use the lower boundary condition, which
simply means $y(u_{min})=0$, to find a relation between the largest
eigenvalue $\lambda$ and $\kappa_{min}$. 
For each $\lambda$, $\kappa_{min}$ is given by the largest zero of the
function $y_2(u)$.

Also in standard notations, the function $y_2(u)$ can be written 
(apart from a constant factor)
$$
y_2(u)=
u^{\frac{1}{2}+i\nu}
U\left(\frac{1}{4}-\nu^2+i\nu,1+i2\nu;u\right),
$$
where $U$ has the following definition:
$$
U(a,b;u)\equiv
\frac{\pi}{\sin \pi b}
\left[
\frac{M(a,b;u)}{(a-b)!(b-1)!} -
\frac{u^{1-b}M(a+1-b,2-b;u)}{(a-1)!(1-b)!}
\right].
$$
The solution $y_2(u)$, having a zero at $u=u_{min}$, is equivalent to requiring
\be
\label{cond}
{\rm Arg}\left[ \tilde M(u_{min}) \right]=-n\pi,
\ee
with $n$ an integer. For very small $u_{min}$ it is sufficient
to include the leading term in an expansion of ${\tilde M}$. In this case
eq. (\ref{cond}) can be approximated by the relation 
$$
\ln u_{min}=-\frac{n\pi}{\nu},
$$
which means that $n=1$ is relevant for the largest zero.
A less severe approximation (for finite $u_{min}$) is given by using 
the first three terms in the expansion of
$M(a,b;u)$ and evaluating ${\rm Arg} [\tilde M(u_{min})]$
numerically. This leads to an approximate relation between $\kappa_{min}$
and the largest eigenvalue $\lambda$, which agrees well with the numerical
solution discussed in appendix \ref{numapp}.

\section{Random walk analogy}
\label{randomwalk}

As mentioned in the main text, for a running coupling constant the chain has similarities with a random walk in a force field. Let us first study a simple example to show what happens if a stochastic
process of the Brownian motion type contains a ``force''. Consider 
a set of discrete ``space-locations'' indexed $1,\ldots,j,\ldots$ 
and assume that at the discrete ``times'' $t=1,\ldots, a, \ldots$ there is
a number of ``objects'', $n(t=a,j)$ at the location indexed $j$. 
Assume further that there is a rule so that the number of objects
changes with time according to
\be
\label{ruleBrown}
n(a+1,j)= \frac{1+\alpha}{2}n(a,j+1)+ \frac{1-\alpha}{2} n(a, j-1)
\ee
Thus there is for a positive $\alpha$ a tendency for the sites with 
a larger space index to lose 
to the sites
with a smaller index  as the time goes by, i.e. the change in the
distribution contain a ``force'' directed towards small space indices.

The equation can (by subtraction of $n(a,j)$ on both sides and a limiting
procedure in space $x$ and time $t$) be rewritten
into (the space-size and time-size parameters are $d$ and $\tau$, respectively)
\be
\label{diffuseBrown}
\frac{\partial n}{\partial t}= \frac{d^2}{2\tau} \frac{\partial^2 n}{\partial
x^2} + \frac{\alpha d}{\tau}\frac{\partial n}{\partial x}
\ee
For $\alpha=0$ we obtain the well-known diffusion equation for $n$, i.e. a
Gaussian with an average and a width both increasing with time $\sim \sqrt{t}$. For $\alpha >0$ 
we find, however, that starting with a narrow distribution at $t=0$, the last term in Eq (\ref{diffuseBrown}) becomes increasingly important when $t$ increases. Thus the time derivative becomes reduced. For $t>\tau/\alpha^2$ the two terms on the right hand side become of equal magnitude in the dominating interval around $x=d\sqrt{t/\tau}$. This implies that the solution approaches the time-independent distribution 
\be
\label{independentx}
n \simeq \exp(-\alpha x/d)
\ee  
This is just what happens for e.g. the atmospheric density of the earth.

This example is not totally a relevant one for the BFKL mechanism because the
occurring
weight distribution  has no simple 
probabilistic interpretation (although
positive definite it is non-normalized as the weight corresponds to a density). Nonetheless there are obvious similarities, all weights contain the notion of a
``direction'', in this case a preferred direction towards small $\kappa$-values
in the cascade chains because of the running coupling, and the result is asymptotically an exponential distribution in $\kappa$.


\begin{thebibliography}{10}
\itemsep -1mm

\bibitem{JSBAGG} B.~Andersson, G.~Gustafson, J.~Samuelsson, \NPB{463}{217}{1996}.

\bibitem{Pinoetal} M.~Ciafaloni, \NPB{269}{49}{1988};\\
  S.~Catani, F.~Fiorani, G.~Marchesini, \PLB{234}{339}{1990},
  \NPB{336}{18}{1990}.

\bibitem{JSHKBAGG} B.~Andersson, G.~Gustafson, H.~Kharraziha and J.~Samuelsson, 
\ZPC{71}{613}{1996}.
 
\bibitem{BFKL} E.A.~Kuraev, L.N.~Lipatov and V.S.~Fadin, \ZETF{72}{373}{1977},
  \JETP{45}{199}{1977}; \\
  Ya.Ya.~Balitsky and L.N.~Lipatov, \YF{28}{1597}{1978}, \SJNP{28}{822}{1978}.

\bibitem{Alanetal} J.~Kwiecinski et al.,\PRD{50}{217}{1995}

\bibitem{DGLAP} V.N.~Gribov, L.N.~Lipatov, \SJNP{15}{438}{1972} and 675;\\
  G.~Altarelli, G.~Parisi, \NPB{126}{298}{1977};\\
  Yu.L.~Dokshitzer, \JETP{46}{641}{1977}.

\bibitem{Yuribook} Yu.L.~Dokshitzer, V.A.~Khoze, A.H.~Mueller
and S.I.~Troyan, Basics of Perturbative QCD (Editions Fronti\'eres,
Gif-sur-Yvette, 1991)

\bibitem{DIPOLE} G.~Gustafson, \PLB{175}{453}{1986};\\
  G.~Gustafson, U.~Petterson, \NPB{306}{746}{1988};\\ B.~Andersson,
  G.~Gustafson, L.~L\"onnblad, \NPB{339}{393}{1990}.

\bibitem{LL} L.~L\"onnblad, \CPC{71}{15}{1992}.

\bibitem{LDCMC} H.~Kharraziha, L.~L\"onnblad, LU-TP 97-21,
 H.~Kharraziha, L.~L\"onnblad, LU-TP 97-34

\bibitem{JETSET} T.~Sj\"ostrand, \CPC{82}{74}{1994}
  
\bibitem{JanChebyshev}J.~Kwiecinski, A.D.~Martin, P.J.~Sutton,
\PRD{52}{1445}{1995}

\bibitem{NUMREC} W.H.~Press, S.A.~Teukolsky, W.T.~Vetterling and B.P.~Flannery,
NUME\-RICAL RECIPES in C {\em The Art of Scientific Computing} 2:nd ed.
(Cambridge University Press).

\bibitem{maths} M.~Abramowitz and I.A.~Stegun, Handbook of Mathematical
Functions (DOVER PUBLICATIONS INC., NEW YORK).

\end{thebibliography}
\end{document}